\documentclass[journal,10pt]{IEEEtran}
\usepackage[utf8]{inputenc}
\usepackage{setspace}
\usepackage{epsfig}    
\usepackage{amsfonts}
\usepackage{amssymb}
\usepackage{amsmath}
\usepackage{comment}
\usepackage{multirow}
\usepackage{psfrag}
\usepackage[linesnumbered,ruled]{algorithm2e}
\usepackage{soul} 
\usepackage{float}
\usepackage{tikz}
\usetikzlibrary{arrows.meta, decorations.markings}
\usetikzlibrary{automata}
\usepackage{lipsum}  
\usepackage{subcaption}
\usepackage{diagbox}
\usepackage{url}
\usepackage{hyperref}
\usepackage{stfloats}
\usepackage{cite}
\usepackage[normalem]{ulem}
\usepackage{tabulary}
\usepackage{tabularx}

\newcommand{\nnp}[0]{\text{95th percentile }}

\definecolor{boristext}{rgb}{0.22, 0.44, 0.88}
\definecolor{boriscomments}{rgb}{0.88, 0.04, 0.04}
\definecolor{boristochange}{rgb}{0.2, 0.8, 0.8}

\begin{document}
\title{Wi-Fi Multi-Link Operation: An Experimental Study of Latency and Throughput }


\author{
Marc~Carrascosa,~
Giovanni~Geraci,~\IEEEmembership{Senior~Member,~IEEE,} 
Edward~Knightly,~\IEEEmembership{Fellow,~IEEE,} and
Boris~Bellalta,~\IEEEmembership{Senior~Member,~IEEE} 
\thanks{M.~Carrascosa-Zamacois, G.~Geraci, and B.~Bellalta are with University Pompeu Fabra, 08018 Barcelona, Spain (\{marc.carrascosa, giovanni.geraci, boris.bellalta\}@upf.edu).}
\thanks{E.~Knightly is with Rice University, Houston TX 77005, USA (knightly@rice.edu).}
\thanks{M. Carrascosa-Zamacois and B. Bellalta were supported by WINDMAL PGC2018-099959-B-I00 and WI-XR PID2021-123995NBI00 (MCIU/AEI/FEDER,UE).}
\thanks{G. Geraci was supported by RTI2018-101040-A-I00, PID2021-123999OB-I00, and a ``Ram\'{o}n y Cajal" Fellowship from the Spanish Research Agency.}
\thanks{Part of the materials presented in this article have been presented at IEEE ICC 2022 \cite{carrascosa2021experimentalPublished}.}
}
\maketitle

\begin{abstract}

In this article, we investigate the real-world capability of the multi-link operation (MLO) framework---one of the key MAC-layer features included in the IEEE 802.11be amendment---by using a large dataset containing 5 GHz spectrum occupancy measurements on multiple channels.
Our results show that when both available links are often busy, as is the case in ultra-dense and crowded scenarios, MLO attains the highest throughput gains over single-link operation (SLO) since it is able to leverage multiple intermittent transmission opportunities.
As for latency, if the two links exhibit statistically the same level of occupancy, MLO can outperform SLO by one order of magnitude. In contrast, in asymmetrically occupied links, MLO can sometimes be detrimental and even increase latency. We study
this somewhat unexpected phenomenon, and find its origins to be packets suboptimally mapped to either link before carrying out the backoff, with the latter likely to be interrupted on the busier link.
We cross validate our study with   real-time traffic generated by a cloud gaming application and quantify MLO's benefits for  latency-sensitive applications.
\end{abstract}


\section{Introduction}

Achieving consistent low delay in Wi-Fi networks is a challenge that has attracted growing interest, motivated by new applications with stringent latency constraints, such as gaming, augmented and virtual reality, industrial automation, and remote healthcare---some requiring response times as low as $1$~ms \cite{adame2021time}. Moreover, Wi-Fi access links have  the potential to be the bottleneck in terms of network delay, accounting for more than 60\% of the Round Trip Time in connections to domestic servers \cite{pei2016wifi}. Indeed, operating in license-exempt bands brings about the need to coexist with other wireless networks, along with the inherent uncertainty as to how many transmission opportunities will be available, and when. One way to mitigate such uncertainty is by  employing multiple radio interfaces for packet transmission. At the time of writing, this approach---termed multi-link operation (MLO)---is one of the main features being proposed and developed for IEEE~ 802.11be~\cite{draft11be,lopez2022multi}, the new amendment that is foreseen to be certified as Wi-Fi 7 \cite{garcia2021ieee, khorov2020current, hoefel2020ieee, yang2020survey}.


\subsection{Motivation}

Through MLO, IEEE 802.11be will target efficient operations in all the available bands, i.e., 2.4, 5, and 6 GHz, for load balancing, multi-band aggregation, and simultaneous downlink/uplink transmission \cite{lopez2019ieee}. In 802.11be, a multi-link device is defined as one with multiple affiliated access points (APs) or stations (STAs), and a single MAC service access point to the above logical link control layer \cite{garcia2021ieee}. Multi-link devices could thus transmit and receive packets at the same time, separate the control and data planes, or transmit delay-sensitive traffic through multiple links to ensure its timely reception \cite{adame2021time,deng2020ieee}. Lastly, while MLO is fully transparent to the upper TCP/IP protocols, they will benefit from the faster and more reliable data communication it enables \cite{draft11be}.

As consensus has not yet been reached on the specific implementation details of MLO, recent works have compared the performance of different variants \cite{song2021performance, yang2019ap}, studied the feasibility of simultaneous transmission and reception \cite{levitsky2020study}, and undertaken the optimization of traffic and resource allocation in MLO \cite{lopez2021ieee,park2021latency,bankov2021use}. 
Besides throughput augmentation, latency reduction has been identified as one of the main endeavors of MLO, with its delay performance being the object of several recent studies \cite{naik2021can, bankov2021use,park2021latency,lacalle2021analysis,schwarzenberg2018quantifying,kondo2020low}. These works---and others---have shown that, in many cases, MLO is capable of enabling new applications whose requirements cannot reliably be supported by conventional single-link operation (SLO).

Notwithstanding the insights provided by these works, the literature currently lacks experimental evidence on what performance gains  MLO can attain over SLO, in what practical scenarios, and under which  channel access methods. This gap prompts us to study, for the first time, the performance of MLO by using  spectrum occupancy measurements and real application traces.


\subsection{Contribution and Summary of Results}

In this paper, we utilize our  over-the-air measurements of spectrum occupancy for the entire 5~GHz band  \cite{barrachina2020wi,barrachina2021wi}\footnote{Freely available in the open source WACA dataset: \url{https://github.com/sergiobarra/WACA\textunderscore WiFiAnalyzer}.} and investigate the throughput and latency
performance of MLO when operating on two links.
Atop these traces, which include scenarios with high AP density and crowded environments and span multiple hours, we develop an emulation tool that fuses a Wi-Fi MLO state machine with the high-resolution spectrum measurements. We feed our MLO state machine with Poisson traffic first, and then validate selected experiments with real-time traffic generated by a cloud gaming application. 
Besides legacy Wi-Fi SLO, we study two  MLO channel access modes defined as follows: \emph{(i)} MLO with Simultaneous Transmit and Receive (MLO-STR), in which both interfaces are available and work independently, and \emph{(ii)} MLO with Non-Simultaneous Transmit and Receive (MLO-NSTR), where both interfaces are available but access to the secondary link is conditioned on the primary also being unoccupied.
While our results confirm the potential latency gains of MLO seen in previous works, the use of real spectrum measurements as well as real traffic traces offers new and otherwise inaccessible insights on MLO. 

Our main 
findings can be summarized as follows: 
\begin{itemize}
    \item We show that an MLO AP with two radio interfaces  achieves throughput higher than the maximum SLO throughput in $53\%$ and $28.5\%$ of the cases by using MLO-STR and MLO-NSTR, respectively. When both links are almost always busy, which may correspond to ultra-dense and crowded scenarios, MLO-STR achieves the highest throughput gains as it is able to leverage the intermittent transmission opportunities over multiple links.
    \item We find that when primary and secondary links have statistically symmetrical occupancy, MLO-STR yields order-of-magnitude \nnp latency benefits over SLO, even in the challenging regime of increasing occupancies and traffic. This is because MLO-STR can utilize either available link, and reduce packet waiting time even when it cannot simultaneously utilize both links. 
    \item In contrast, we surprisingly discover that when using two links with asymmetrical occupancy, MLO-STR can sometimes underperform SLO by up to 112\% in terms of \nnp latency. This is owed to packets being suboptimally assigned to an interface before carrying out the backoff, with the latter likely to be interrupted on the busier link. This phenomenon is exacerbated when the asymmetry in channel occupancy increases.
    \item To overcome the aforementioned phenomenon, we define a third MLO channel access approach, denoted MLO-STR+, that employs parallel backoff instances for each  interface and allocates  packets to the interface whose backoff expires first. While MLO-STR+ is a minor variation on MLO-STR, we study it to better understand the design space and ultimate capabilities of MLO.
    \item We further validate our results by using, in addition to real-world channel occupancy measurements,  real-time traffic generated by a cloud gaming application. This final set of experiments confirm our previous findings, and further demonstrate MLO's capabilities to enable  latency-sensitive applications whose traffic load cannot otherwise be delivered in a timely manner through SLO.

    \item Using channel bonding, we show that splitting the channel bandwidth (80 MHz) between two independent MLO links (2x40 MHz) leads to lower delays than using the entire bandwidth for a single link. Further, we show that by using wider links, we have another degree of freedom in the primary  channel used, and correct selection can lead to an 89.5\% delay reduction for MLO-STR+. 
\end{itemize}

The rest of the document is organized as follows. Section~\ref{evaluationmethodology} details the experimental setup, including the dataset and methodologies used. Section~\ref{Throughput} studies the achievable MLO throughput based on the links occupancy. Section~\ref{delay} analyses the MLO delay in symmetrically and asymmetrically occupied links. Section~\ref{decomposition}  introduces MLO-STR+ as a way to improve the observed drawbacks of MLO-STR operation. In Section~\ref{sec:GoogleStadia},  real traffic traces are used to validate previous results obtained using Poisson traffic. Section \ref{bonding} studies the use of channel bonding coupled with MLO. Section \ref{RW} includes the related work, and finally, Section \ref{conc} concludes the paper.

\section{MLO Protocol and Experimental Setup}\label{evaluationmethodology}


We consider a scenario consisting of multiple Basic Service Sets (BSS) and STAs accessing multiple channels in the 5 GHz band. We use a dataset containing real spectrum information from a crowded stadium. Since the dataset contains the aggregate signal strength received at the endpoint, we cannot identify individual devices and we focus on a single AP and STA pair equipped with two MLO-capable interfaces each, denoted the MLO-BSS. We do not add extra model/simulation-driven sources of contention so as to make all our findings only dependent on the real traces. 

\begin{figure}[t!!!]
    \centering
    \includegraphics[width= 0.9\columnwidth]{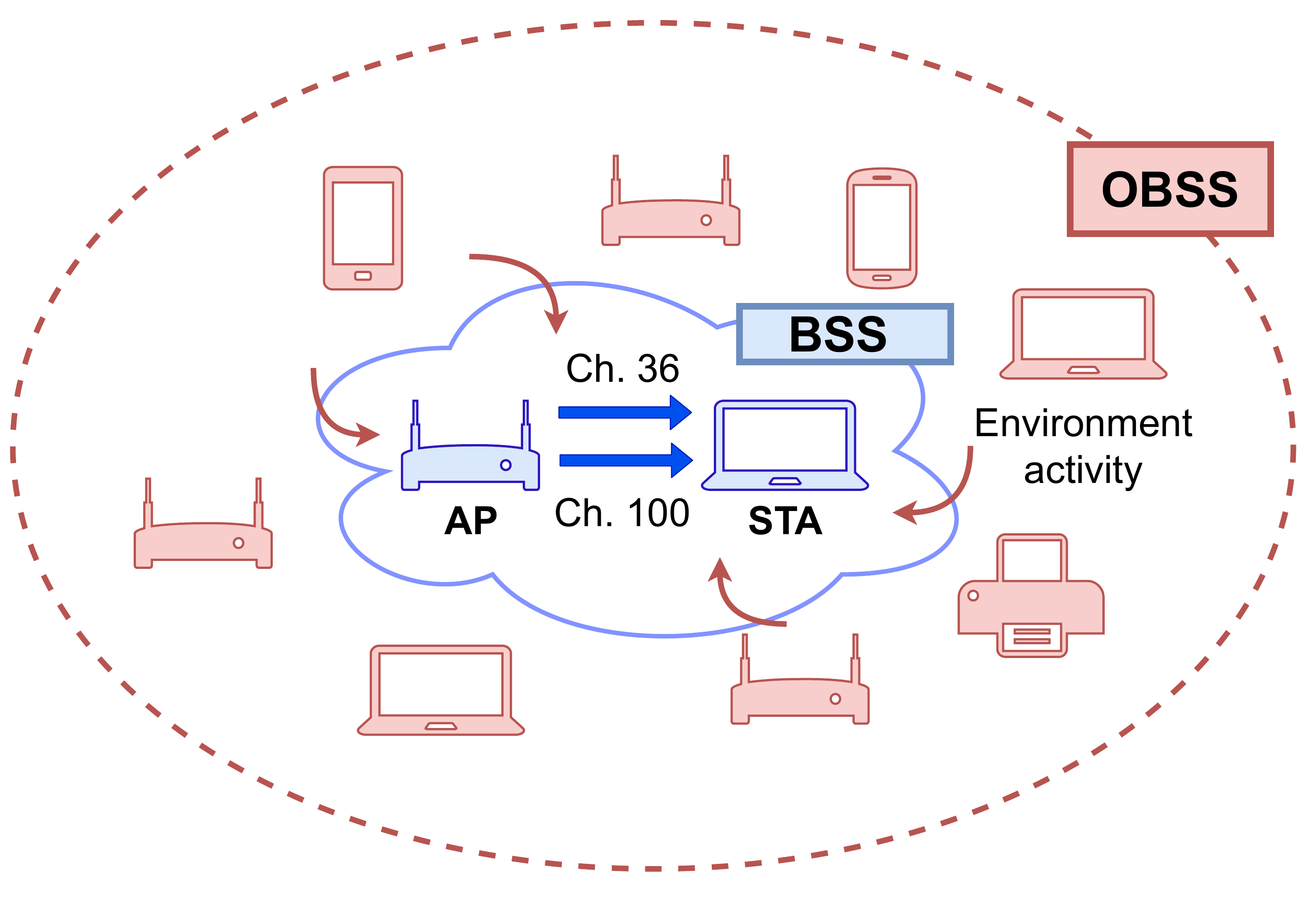}
    \caption{Scenario considered. The WACA dataset is used to characterize the environment activity (red) observed by the target BSS (blue) on channels 36 and 100 in the 5 GHz band. }
    \label{fig:system_model}
\end{figure}

The MLO interfaces operate in the 5~GHz band on 20~MHz-wide channels, respectively on channel 36 (low 5 GHz) and channel 100 (high 5 GHz), which we denote as \emph{primary} and \emph{secondary}, respectively. On both channels, the MLO-BSS observes the environment activity, i.e., the transmissions generated by the Orthogonal Basic Service Sets (OBSS). The MLO-BSS and OBSS under consideration are illustrated in Figure~\ref{fig:system_model} in blue and red, respectively. The environment activity from the OBSS is characterized by the WACA dataset (Section \ref{wacadesc}).
For the MLO-BSS, we only consider downlink traffic, i.e., from the AP to the STA. We initially assume packet arrivals to follow a Poisson process, and transmitted packets to have a constant size of $L=12000$ bits. In Section~\ref{sec:GoogleStadia}, we consider real-time data traffic instead. Since we do not have client-AP SNR traces, any MCS variation would need to be purely model driven. We therefore employ a fixed MCS (256-QAM, 5/6) in all links.

Next, we detail the channel access schemes considered, the measurement-based channel occupancy model, and the performance evaluation methodology.

\subsection{Multi-link Channel Access Policies}

We study three channel access policies for the MLO-BSS, namely:
\begin{itemize}
    \item Conventional single-link operation (SLO), where only the primary interface is available.
    \item Multi-link operation with Simultaneous Transmit and Receive (MLO-STR), where both interfaces are available and work independently.
    \item Multi-link operation with Non-Simultaneous Transmit and Receive (MLO-NSTR), where both interfaces are available but access to the secondary is subjected to the state of the primary link. 
\end{itemize}

In particular, the two MLO channel access schemes operate as follows:
\subsubsection{MLO with Simultaneous Transmit and Receive (MLO-STR)} The two radio interfaces operate independently and asynchronously. The first packet waiting for transmission in the buffer is allocated to the first radio interface that becomes available. If both radio interfaces are available, the packet is randomly allocated to either. Once a packet is allocated to an interface, it starts the channel access procedure by initializing a backoff instance.
\subsubsection{MLO with Non-Simultaneous Transmit and Receive (MLO-NSTR)} One radio interface always acts as primary, and the other always as secondary. When there are packets waiting for transmission, the primary interface undergoes contention to access the channel. Once the backoff counter reaches zero, packets are sent through the two interfaces if the secondary one has been idle for at least a PIFS interval before the backoff expiration. Otherwise, only a single packet is transmitted through the primary link.

\begin{figure*}[ht!!!]
    \centering
    \includegraphics[width= 0.9\textwidth]{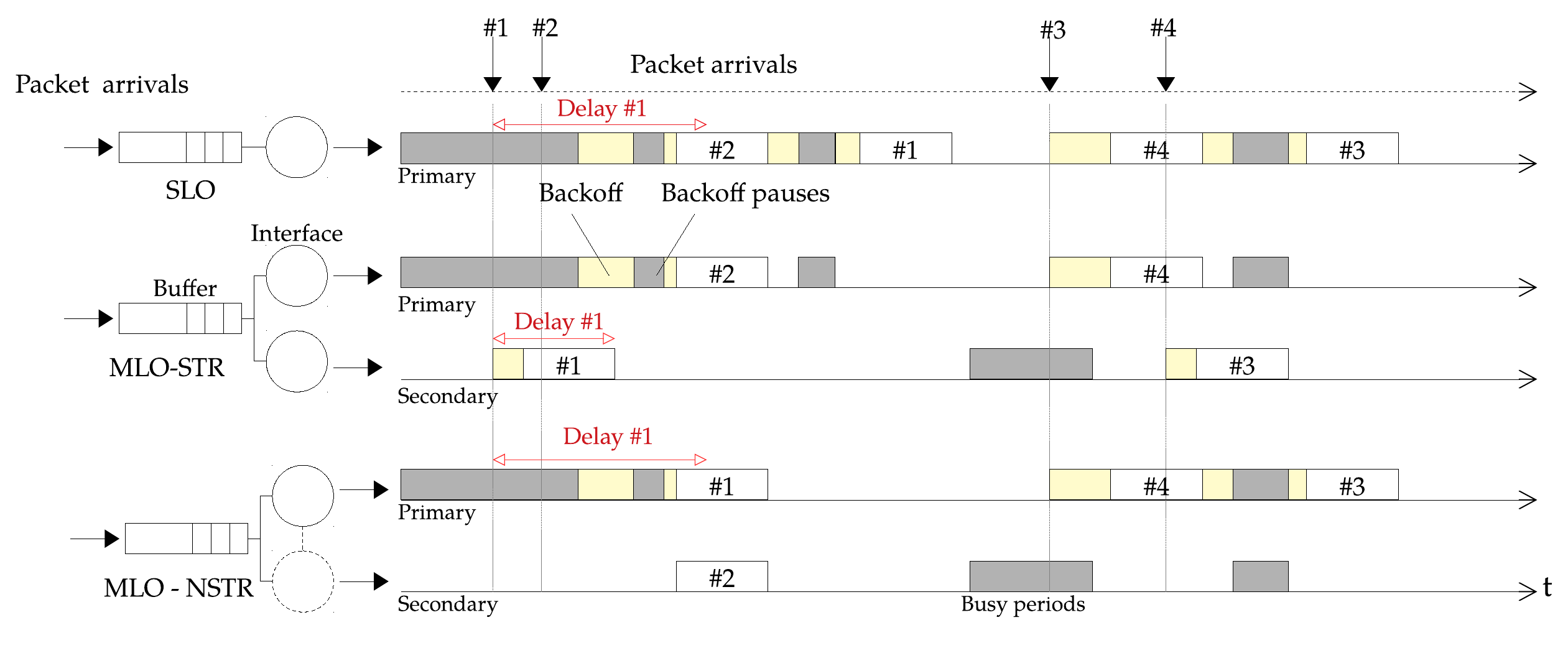}
    \caption{Illustration of SLO, MLO-STR, and MLO-NSTR operations. Grey, yellow, and white bars respectively indicate occupied channels, random backoffs, and packet transmissions. Packet transmissions include both the data part and the corresponding ACK, as well as DIFS and SIFS.}
    \label{Fig:MLO_channelaccess}
\end{figure*}

Figure \ref{Fig:MLO_channelaccess} exemplifies SLO, MLO-STR, and MLO-NSTR operation. SLO follows default Wi-Fi access, where packets are sequentially transmitted over a single link, with packet 1 being the first to be transmitted in the timeline before starting backoff for packet 2, and so on. In the case of MLO-STR, arriving packets are allocated to whichever interface becomes available first. This results in a significant delay reduction for packets \#1, \#2, and \#4. In the case of MLO-NSTR, the secondary link’s dependence on the primary sometimes prevents using the two radio interfaces efficiently. As a result, and unlike MLO-STR, the delay for packets \#1 and \#4 cannot be reduced with respect to SLO.

In order to evaluate the above MLO schemes, we  extended the IEEE 802.11 state machine originally developed in \cite{barrachina2021wi} by adding  functionalities to accurately reproduce the temporal system dynamics under finite traffic loads (i.e., non-full buffer conditions). In order to isolate the combined effect of the access scheme and channel occupancy, we assume a fixed modulation and coding scheme on both interfaces -a 256-QAM with coding rate 5/6 and 2 spatial streams- yielding a transmission time of $0.172$~ms (DATA+SIFS+ACK). Since our FCB-WACA dataset (described next) contains measurements taken with a periodicity of 10~$\mu$s, we have rounded the duration of IEEE 802.11 timings to integer multiples of $10~\mu$s, setting the duration of a backoff empty slot, SIFS, and DIFS to $10~\mu$s, $10~\mu$s, and $30~\mu$s, respectively. Such small approximation implies no loss of generality, as discussed in~\cite{barrachina2021wi}. The value of the CW$_{\min}$ used in all cases is $15$.


\subsection{Measurement-based Channel Occupancy Dataset}\label{wacadesc}

To investigate the performance of the MLO-BSS in a real-world setting---i.e., while considering OBSS activity---we employ the \emph{WACA dataset}, containing over-the-air measurements of the 5 GHz band occupancy that we have recently collected and made publicly available \cite{barrachina2020wi,barrachina2021wi}. This dataset was obtained by conducting extensive measurement campaigns on different days and in multiple locations, including a sold-out football stadium (F.~C. Barcelona's Camp Nou). In this paper, we focus only on the football stadium measurements since they range from completely idle to fully occupied channels. We will refer to such subset of measurements as the \emph{FCB-WACA dataset}.

\begin{figure}[t!!]
    \centering
    \includegraphics[width=\columnwidth]{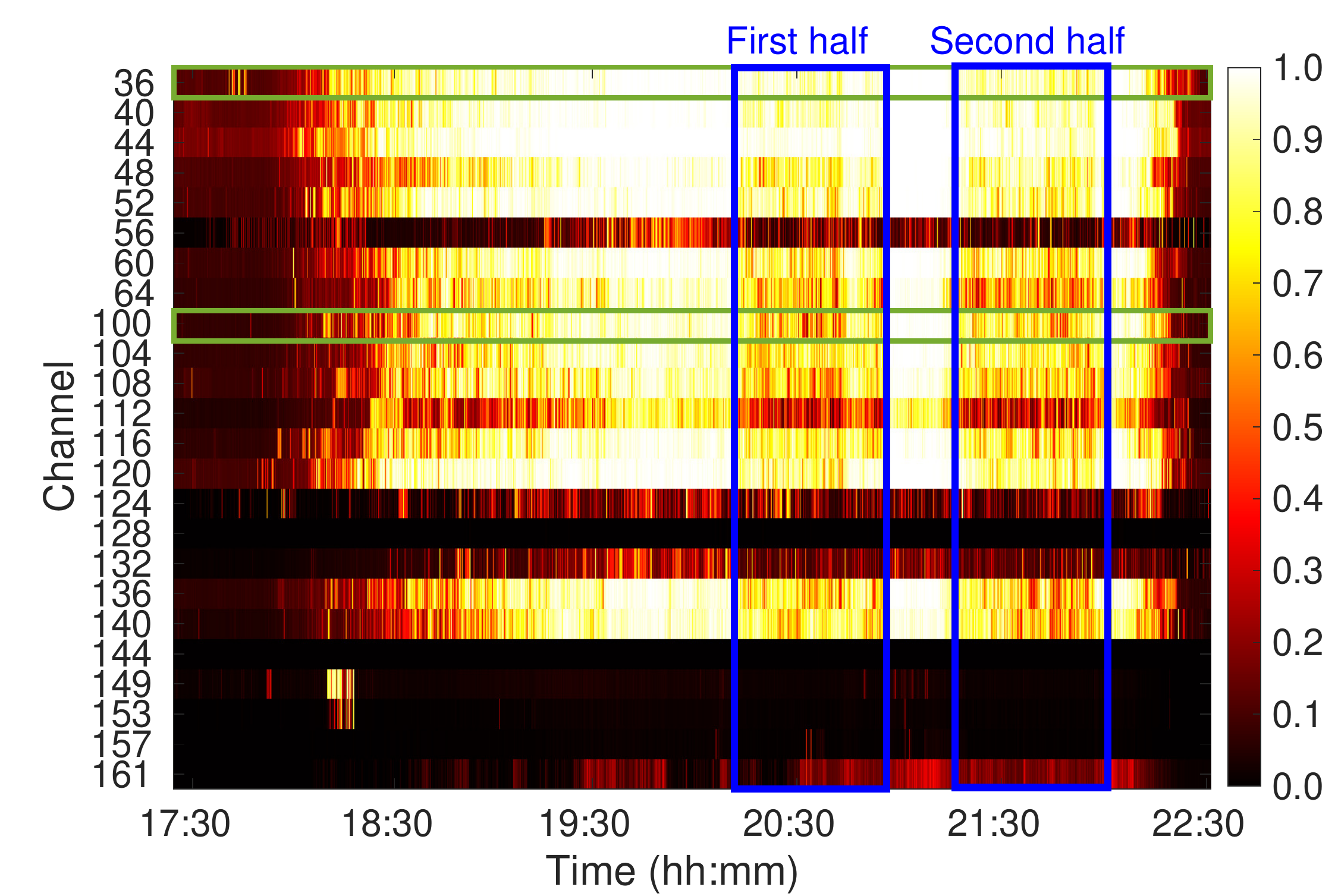}
    \caption{Average channel occupancy in the FCB-WACA dataset. Channels 36 and 100, used in our experiments, are highlighted horizontally. Time intervals corresponding to the first- and second-half of the football game are highlighted vertically.}
    \label{f3}
\end{figure}

The FCB-WACA dataset spans 5 hours and contains 2000 samples of the Received Signal Strength Indicator (RSSI) for each of the 24 20-MHz channels in the 5 GHz band.\footnote{The F.~C. Barcelona Camp Nou's network only supports 20 MHz channels without channel bonding.} Each sample lasts one second and consists of 1000 consecutive 10~$\mu$s measurements containing the aggregate signal strength of all nodes in the area. In Figure~\ref{f3}, the spectrum occupancy in the FCB-WACA dataset is displayed as the average number of busy slots in each one-second sample, with a slot considered busy if its RSSI is above -83.5 dBm. We note from Figure~\ref{f3} that the channel occupancy varies across the measurement campaign, exhibiting: (i) predominantly empty channels prior to the football game, (ii) increasing occupancy up until the game starts and during half-time recess, (iii) lower occupancy during the first- and second-half, and (iv) a rapidly decreasing occupancy from the end of the game onwards.

\begin{figure}[t]
    \centering
    \includegraphics[width=0.4\textwidth]{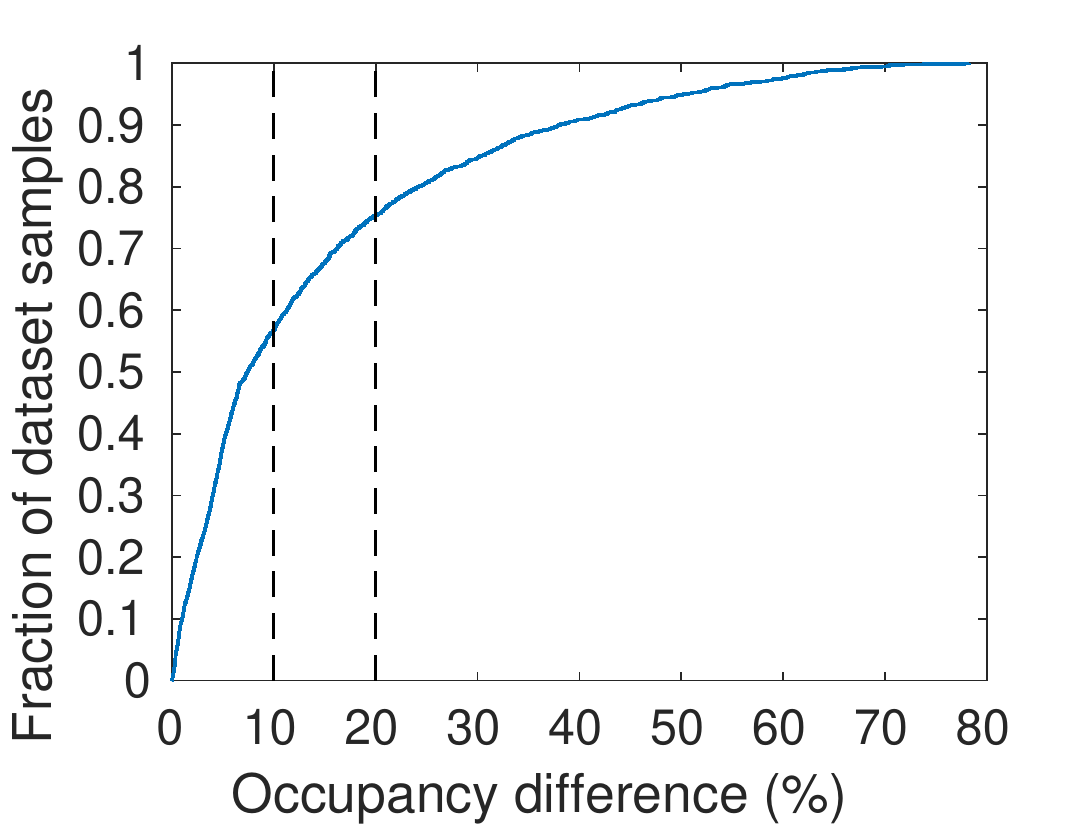}
    \caption{Distribution of the difference in occupancy between channels 36 and 100.}
    \label{symmetry}
\end{figure}

While the temporal evolution of channels 36 and 100 appears similar from Figure \ref{f3} at the macroscopic level, the same does not necessarily hold when observing concurrent one-second samples from the two channels and comparing their average occupancy. To quantify the occupancy disparity, Figure~\ref{symmetry} shows the distribution of the absolute value of the difference in occupancy between the two channels. Although such difference is lower than 10\% (resp. 20\%) in 75\% (resp. 57\%) of the samples, there is also a non-negligible number of cases with high occupancy disparity. The latter prompts us to evaluate the performance of MLO channel access schemes both under symmetric and asymmetric channel occupancy, as discussed in the remainder of the paper. 

In what follows, we employ the FCB-WACA dataset to investigate how different combinations of primary and secondary channel occupancies affect the MLO-BSS performance. In particular, we assume that the MLO-BSS perceives the same spectrum activity as the one captured in the FCB-WACA dataset, and it contends for channel access accordingly. To address the interaction between the simulated node and the traces, we implement the same ``hinder" interaction model as in~\cite{barrachina2021wi}. Using the hinder model, once the MLO-BSS starts a transmission, we assume that the OBSS devices are able to sense the on-going transmission and defer accordingly, allowing the MLO-BSS transmission to finish successfully, thus avoiding collisions. As it is shown in \cite{barrachina2021wi}, the hinder interaction model keeps the same implicit channel access fairness as in CSMA.

\subsection{Trace-based Simulation Methodology}\label{tracebasedsimulationmethodology}

\begin{figure}[tb]
    \centering
    \includegraphics[width = 0.45\textwidth]{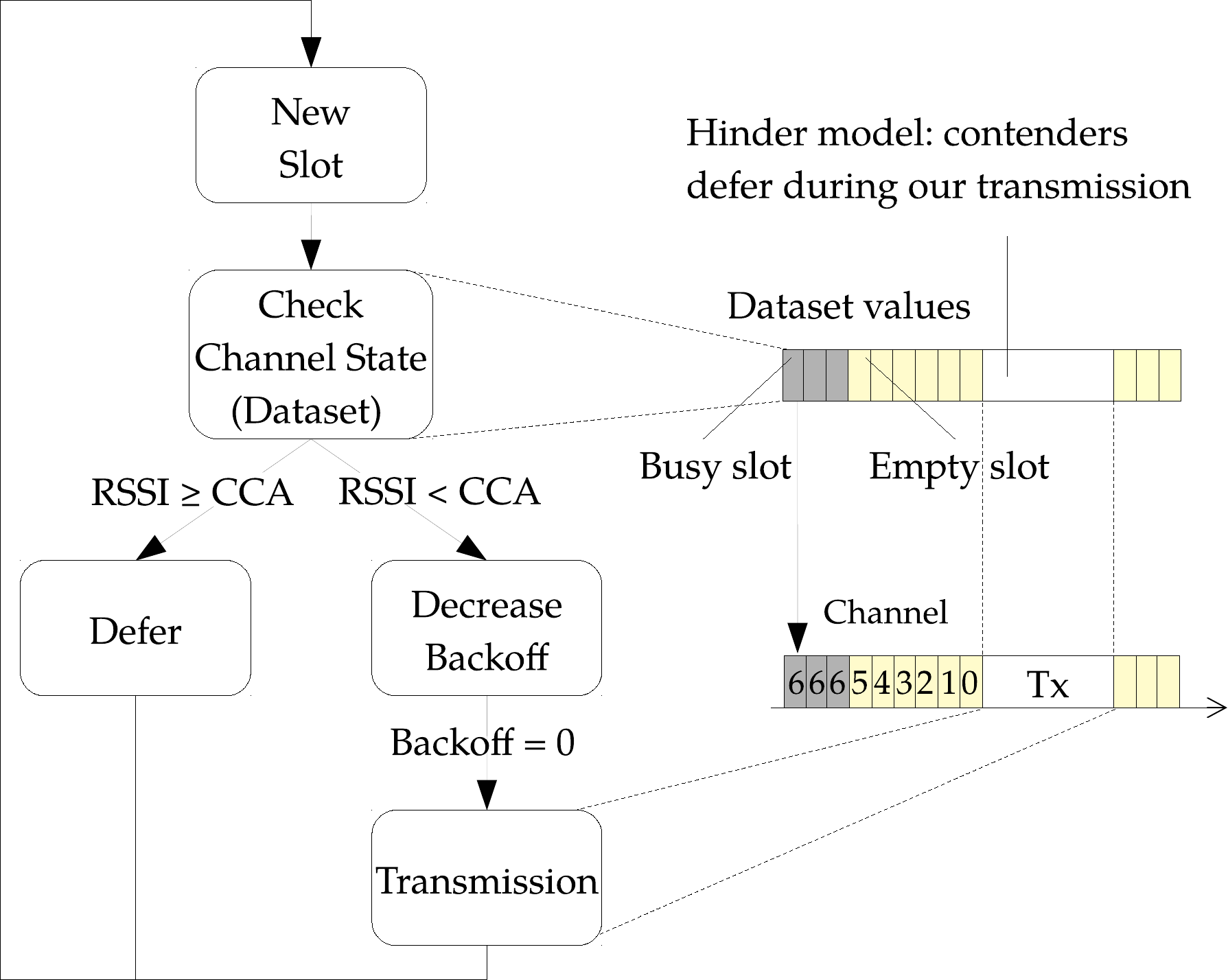}
    \caption{Wi-Fi state machine (per interface) using WACA dataset to determine channel state. }
    \label{integration}
\end{figure}

Figure \ref{integration} illustrates how the WACA dataset is used by the Wi-Fi state machine in each of its interfaces. As previously stated, we accurately follow the WACA dataset occupancy measurements to determine the MLO-BSS channel access dynamics.

\begin{figure}[ht]
    \centering
    \includegraphics[width = \columnwidth]{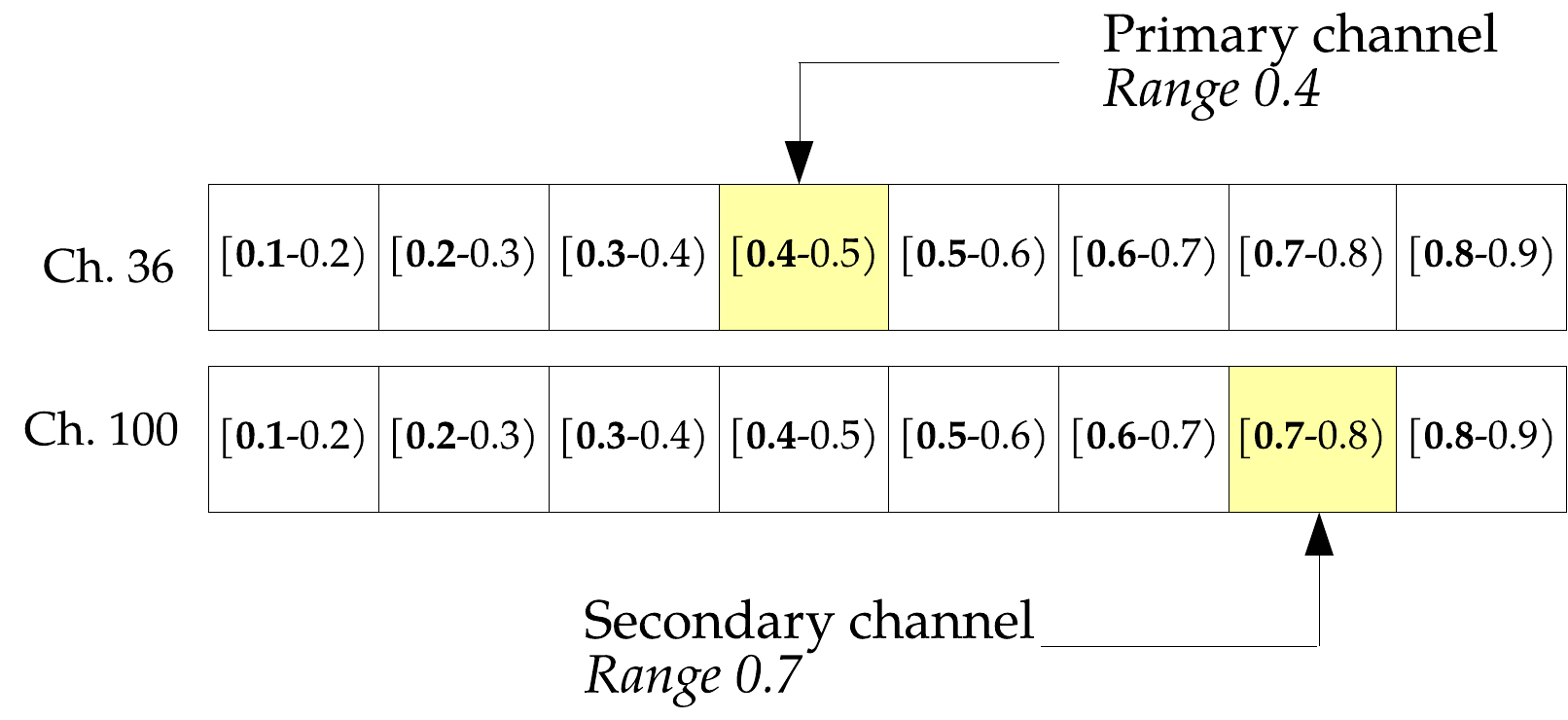}
    \caption{Partitioning the available traces in the dataset into different average channel occupancy regimes.}
    \label{Fig:MLO_Dataset}
\end{figure}

In order to study the effect of channel occupancy on latency, we treat both channels as independent and partition the available traces in our dataset into different average channel occupancy regimes: \{10\%, 20\%, \ldots, 90\%\}, as illustrated in Figure \ref{Fig:MLO_Dataset}. 

We perform $20$ experiments for each combination of channel occupancy and traffic load, with each experiment considering a pair of one-second spectrum samples. Samples from channel 36 are assigned to the primary MLO link, and channel 100 to the secondary link. 

Each experiment is carried out as follows:
\emph{(i)} We select the occupancy regime of interest for the primary and secondary links, e.g., 10\% and 40\%, respectively.
\emph{(ii)} We combine uniformly at random one spectrum sample each for the primary and secondary links from the dataset.
\emph{(iii)}  For each spectrum sample pair and given a particular traffic load of interest, we compute the packet arrival times at the AP.
\emph{(iv)}  We execute the Wi-Fi state machine for SLO, MLO-STR, and MLO-NSTR access policies. The same packet arrival times are considered in all cases to allow a direct comparison. 
\emph{(v)} We store the individual delay experienced by each packet in each experiment.

We then combine the per-packet results (i.e., the individual packet delays) from all runs to obtain the average and the 95th percentile delay. We
guarantee that all results are obtained under stability conditions (i.e., the AP does not become backlogged) and thus we discard any experiment where less than 95\% of all the transmitted packets are
received.

\section{Origins of Throughput Gains}\label{Throughput}

\begin{figure}[ht!]
\centering
\begin{subfigure}[b]{0.45\textwidth}
    \includegraphics[width = \textwidth]{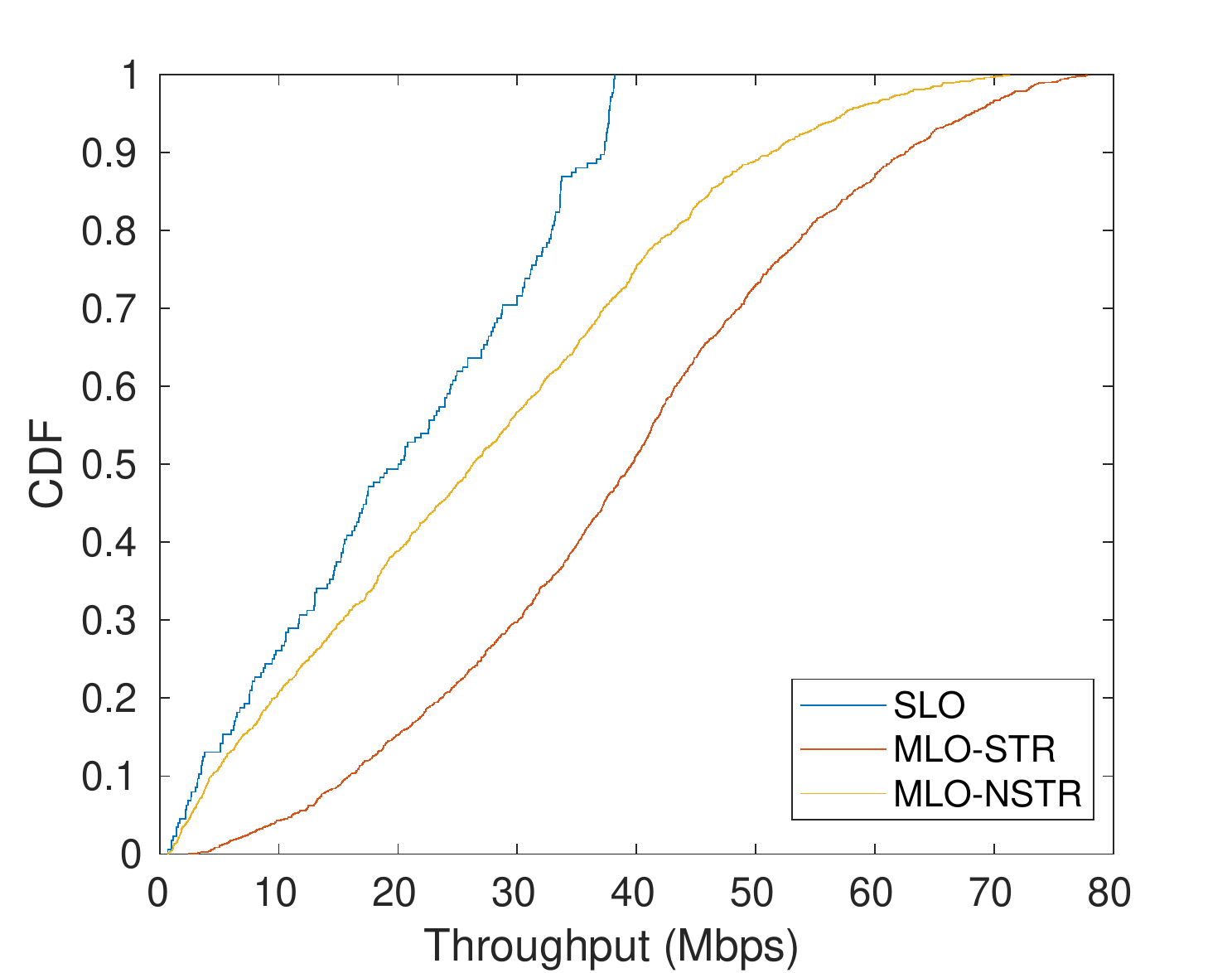}
    \caption{Cumulative distribution function of the throughput achieved by each of the three channel access modes.}
    \label{cdf}
\end{subfigure}

\begin{subfigure}[b]{0.45\textwidth}
    \includegraphics[width = \textwidth]{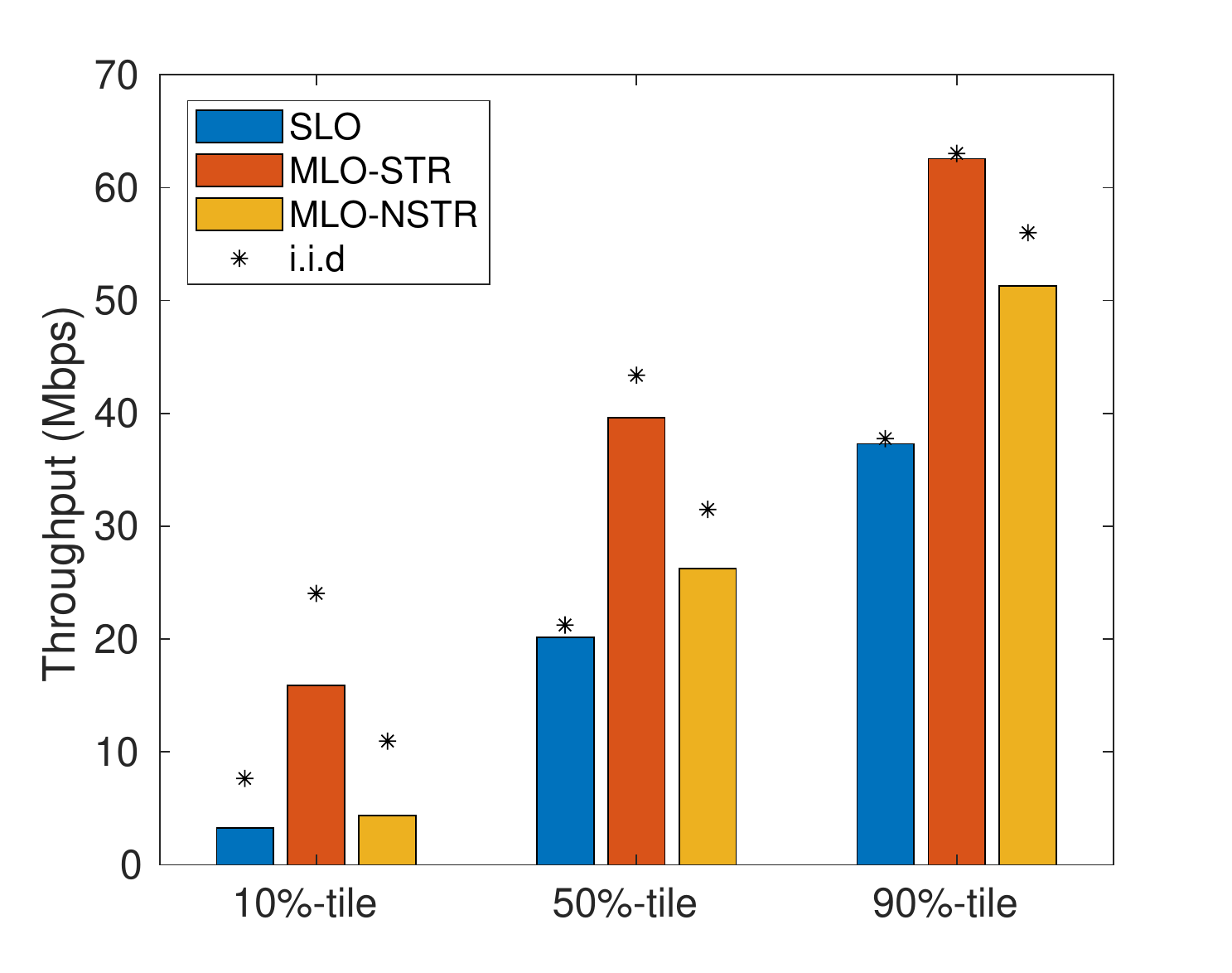}
    \caption{Throughput for each channel access method in terms of 10\%, 50\%, and 90\%-tile using our trace-based simulator (bars). Values obtained analytically via an i.i.d model (stars) are also shown for comparison.}
    \label{iid}
\end{subfigure}
\caption{Throughput statistics for each of the three channel access methods considered.}
\label{ratiosb}
\end{figure}

\begin{figure*}[ht!]
\centering

\begin{subfigure}[b]{0.4\textwidth}
    \includegraphics[width = \textwidth]{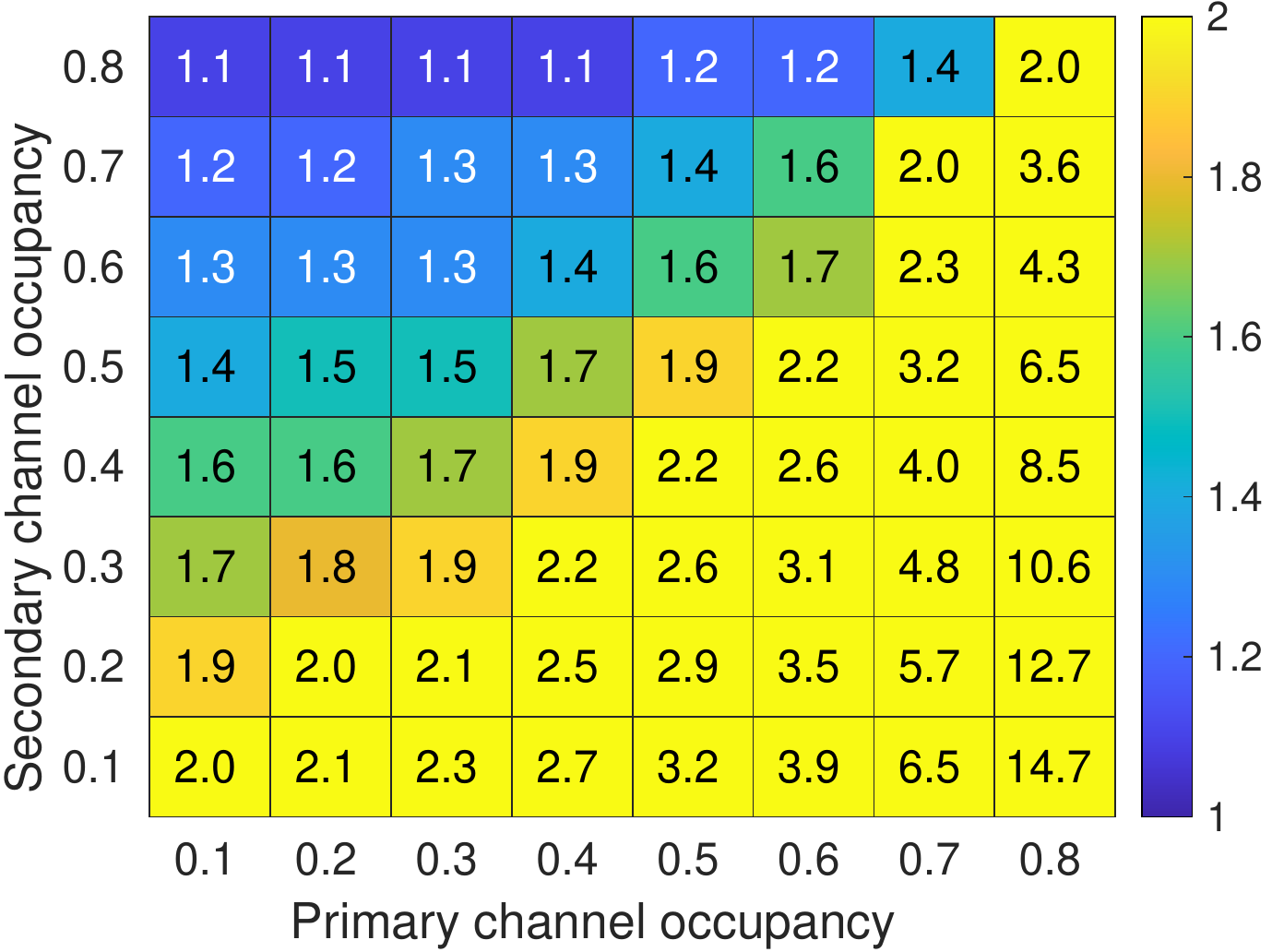}
    \caption{Throughput gain of MLO-STR}
    \label{asynch}
\end{subfigure}
\begin{subfigure}[b]{0.4\textwidth}
    \includegraphics[width = \textwidth]{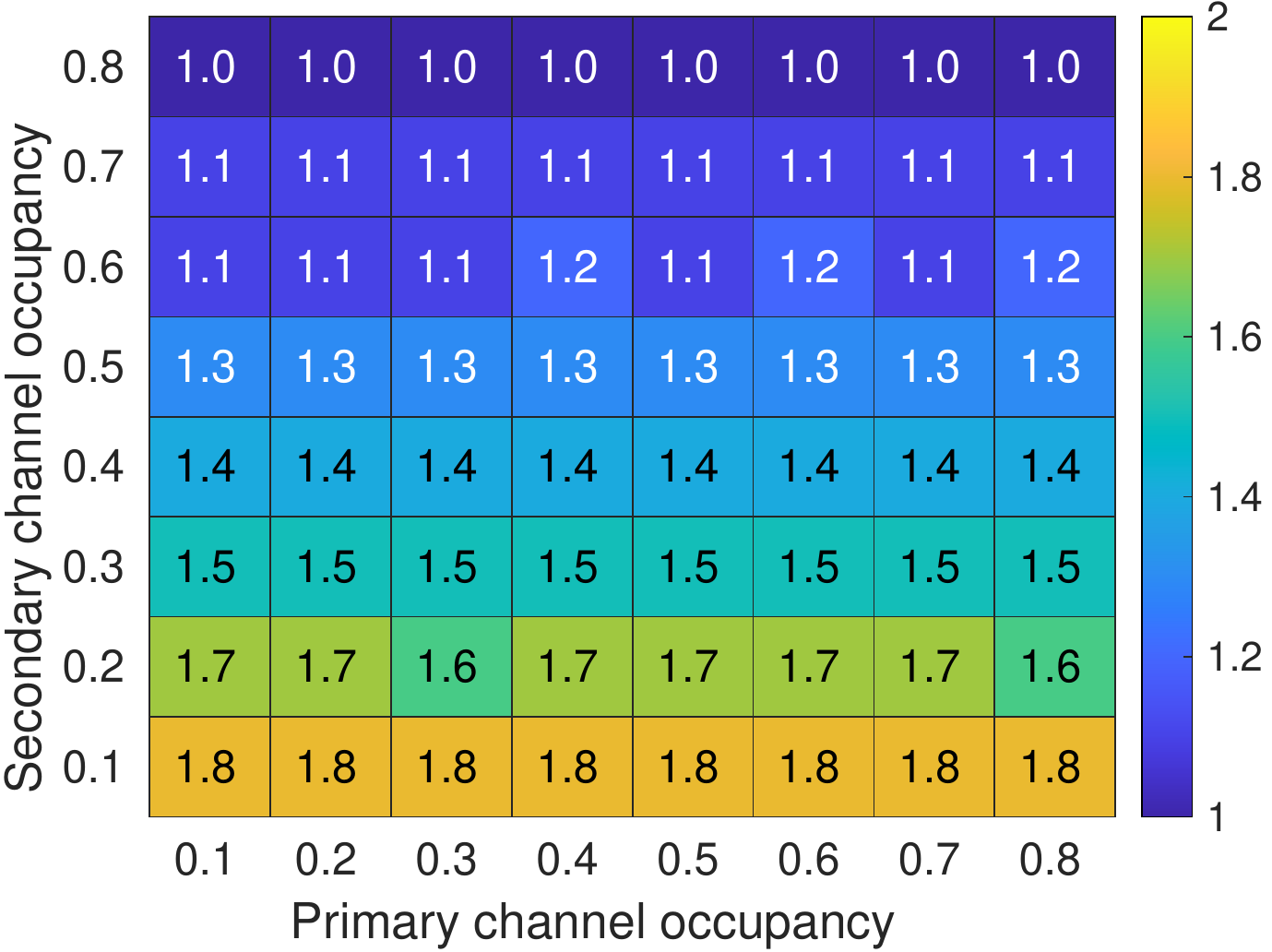}
    \caption{Throughput gain of MLO-NSTR}
    \label{synch}
\end{subfigure}
\caption{Average MLO throughput normalized to SLO. The upper bound of the colorbar is set to 2 to highlight ratios between 1 and 2.}
\label{ratios}
\end{figure*}
 
Owing to the availability of an extra radio interface at the AP, MLO is guaranteed to increase the BSS throughput with respect to single-link operation. However, the effectiveness of MLO in practice, including challenging scenarios such as  crowded and rapidly changing environments, is tied to the instantaneous occupancy patterns of the links in use. In this section, we study this aspect by analyzing the effectiveness of each MLO policy in terms of the throughput gains attainable over a SLO baseline. Aiming for a general understanding, we consider all possible combinations for the statistical occupancy of the two available links, and study how the latter affects the attainable gains. In this section we consider that the AP is fully backlogged.


\subsection{Throughput Distribution}

MLO-STR operates interfaces independently, thus aggregating the available airtime from different links. MLO-NSTR however, synchronizes its secondary interface to the primary. One can thus expect MLO-NSTR to require links with similar activity patterns to achieve additive capacity. However, activity patterns on different links are likely to be completely independent due to the number of contenders, their traffic loads, and random access mechanisms.

Figure~\ref{cdf} shows the empirical CDF of the throughput achieved with each channel access mode across all channel occupancy combinations for our trace-based simulation. The SLO throughput increases proportionally to the airtime available, with the lowest value of $0.672$ Mbps corresponding to the highest occupancy of 0.8, and the highest value of $38.1$ Mbps to the lowest occupancy of 0.1. MLO-STR and MLO-NSTR have a different throughput distribution as a result of using two links, yet they function similarly. 
MLO-STR leverages the secondary link independently from the primary and thus offers a throughput improvement over SLO across the whole CDF, showing a remarkable $5\times$ gain in terms of $1\%$-worst throughput. MLO-NSTR behaves similarly to SLO in the lower tail, when it struggles to find simultaneous transmission opportunities on both links, and similarly to MLO-STR in the upper tail, when the low occupancy of the primary link allows frequent simultaneous transmissions.

We next evaluate the impact of time and channel statistical dependence on throughput achievable by the two MLO policies. To do so, we compare the throughput values obtained from the original traces against the those provided by a simple baseline model, the latter built under the assumption that the temporal activity of each channel is independent and identically distributed (i.i.d.), with an average occupancy value $\rho$ matched from data.

Under such a baseline model, the mean throughput for the three access modes can be approximated as follows:
\begin{itemize}
    \item SLO throughput is given by
        $\mathrm{Th}_{\rm SLO}=(1-\rho_1)L/T$, 
    where $\rho_1$ is the occupancy of the primary link, $L$ is the packet size (12,000 bits) and {$T$ ($0.277$~ms) is the packet transmission time ($0.172$~ms) plus a single DIFS and an average backoff duration assuming the link is sensed free all time (DIFS+$\frac{\text{CW}_{\min}}{2}10~\mu$s+DATA+SIFS+ACK).}
    \item MLO-STR throughput is computed as
        $\mathrm{Th}_{\rm MLO-STR}= (2-\rho_1-\rho_2)L/T$, 
        
    where $\rho_2$ is the occupancy of the secondary link.
    \item MLO-NSTR throughput is computed as
        $\mathrm{Th}_{\rm MLO-STR}= (1-\rho_1)(2-\rho_2)L/T$,
\end{itemize}

Figure~\ref{iid} shows the throughput achieved by both our trace-based simulation and our i.i.d. model in terms of 10\%, 50\% and 90\%-tile. For the 90\%-tile, our baseline model matches the results closely, while for the 50\% and 10\%-tile the model predicts higher throughput. These differences for the 50\% and 10\%-tiles are explained by the frequent backoff interruptions that appear in our trace-based simulation when the links occupancy increases, an aspect not captured by the simple model. In the case of MLO-NSTR, in addition to the previous effect, the observed differences show that the model is also optimistic regarding the probability to find the two links idle since it does not either capture the existing temporal correlation ---even if low--- between links. 

\textit{Findings:} While MLO-STR is able to leverage the independent occupancy of multiple links to achieve better than additive capacity, MLO-NSTR only reaches near-additive capacity when the two links are almost completely idle, making it less suitable for crowded scenarios.

\subsection{Throughput Gains and Spectrum Occupancy}
\label{sec:lo}

The benefits of MLO compared to SLO are strongly affected by  spectrum occupancy. Importantly, the occupancy of the primary and secondary link affect the MLO state machine quite differently as described in Section \ref{evaluationmethodology}. In light of the above, we conduct separate experiments according to the average per-link occupancy observed on each 1-second sample (as described in Section \ref{tracebasedsimulationmethodology}).

Figure~\ref{ratios} shows the throughput of the two MLO modes normalized to the one attained by SLO, depicted as both numerical values and a heat map. First, we consider 
Figure~\ref{asynch}  and MLO-STR opportunistic and asynchronous access. On the diagonal, both links have nearly identical occupancy and the gain compared to SLO is close to twofold, as expected  since the two links are accessed independently.
In contrast, when the secondary link has lower occupancy than the primary (bottom right of Figure~\ref{asynch}) the throughput gain of MLO-STR vs. SLO is highest, with a maximum factor of $14.7$ when the primary and secondary links are busy $80\%$ and $10\%$ of the time, respectively. While the absolute throughput values are omitted from the figure, for this particular case MLO-STR achieves $40.5$ Mbps, while SLO achieves only $2.76$ Mbps.

Figure \ref{synch} depicts the results obtained for MLO-NSTR. Compared to MLO-STR, the gains are dramatically reduced and have a maximum of $1.8$, due to the requirement that the primary link be unoccupied. In terms of absolute throughput values, when the primary and secondary links are respectively occupied $10\%$ and $80\%$ of the time, MLO-NSTR achieves a mean throughput of $39$ Mbps. This value however plummets to a mere $5$ Mbps if the primary/secondary link occupancy is reversed.
That is, a busy primary link prevents MLO-NSTR from achieving high throughput, even when availing of an idle secondary link. 

\textit{Findings:} MLO-STR's independent link access can yield over $14\times$ throughput gains compared to SLO by overcoming a densely occupied link and taking advantage of a secondary sparse link. Conversely, the performance of MLO-NSTR is tied to the occupancy of the primary link, as the second link can only be accessed when the primary is available too. As a result, its throughput gain over SLO is at most twofold.


\section{Delay Experiments}\label{delay}

A key performance objective of MLO is to improve delay performance, including both  average and tail performance such as the delay distribution's 95th percentile. As the above throughput experiments demonstrated the critical role of link occupancy, we partition the study into two cases. First, we study the delay under the assumption that the two links used have similar occupancy, showing how both MLO access modes deal with a symmetrical increase in the occupancy of the two links and diminishing transmission opportunities. Then, we consider a secondary link statistically more occupied than the primary, and study how each MLO mode adapts to a reduced availability of its secondary link.  

\subsection{Symmetrically Occupied links}

\begin{figure*}[ht]
\centering
\begin{subfigure}[b]{0.325\textwidth}
    \includegraphics[width = \textwidth]{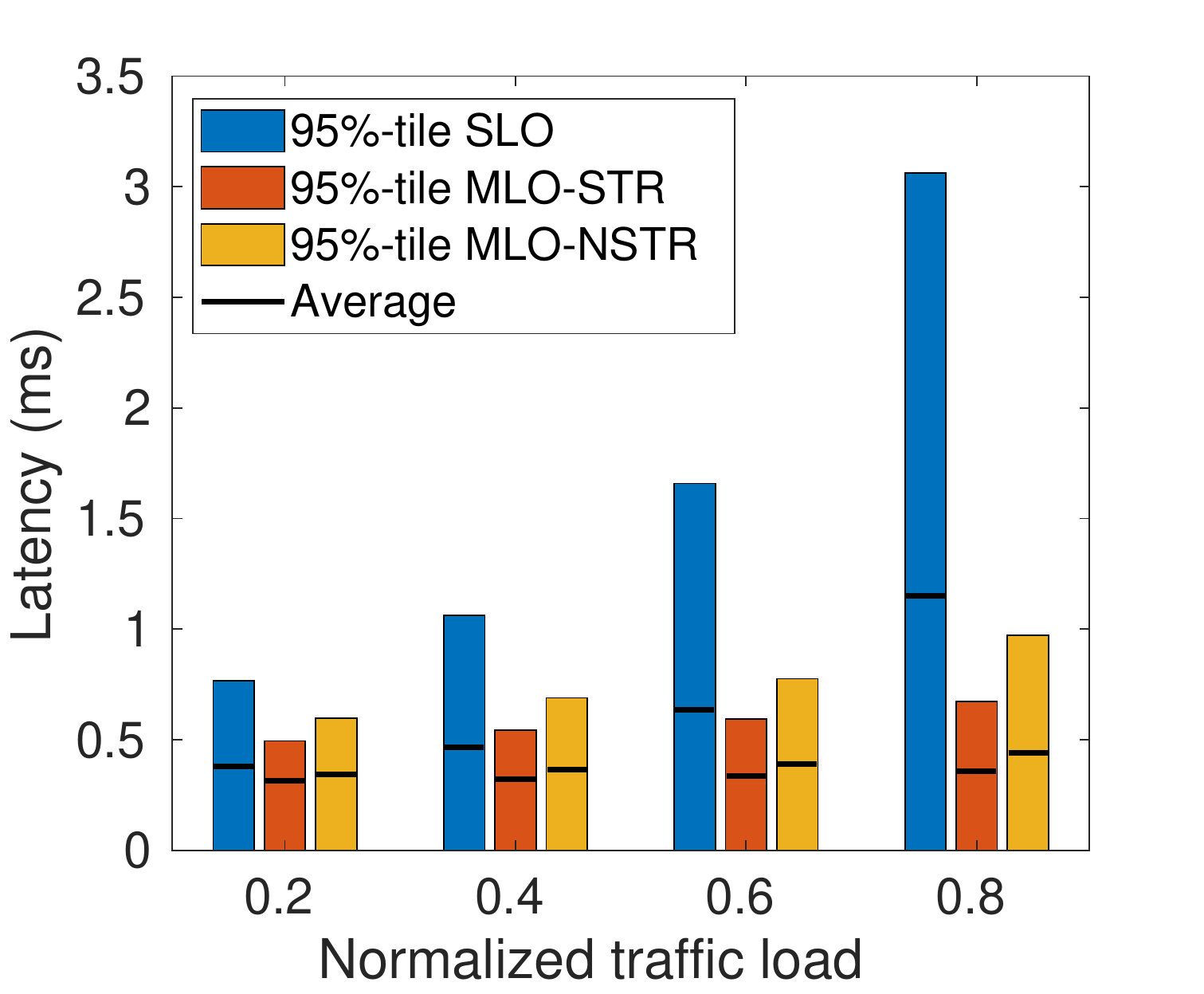}
    \caption{ 10\% occupancy on both links   }
    \label{sym1}
\end{subfigure}
\begin{subfigure}[b]{0.325\textwidth}
    \includegraphics[width = \textwidth]{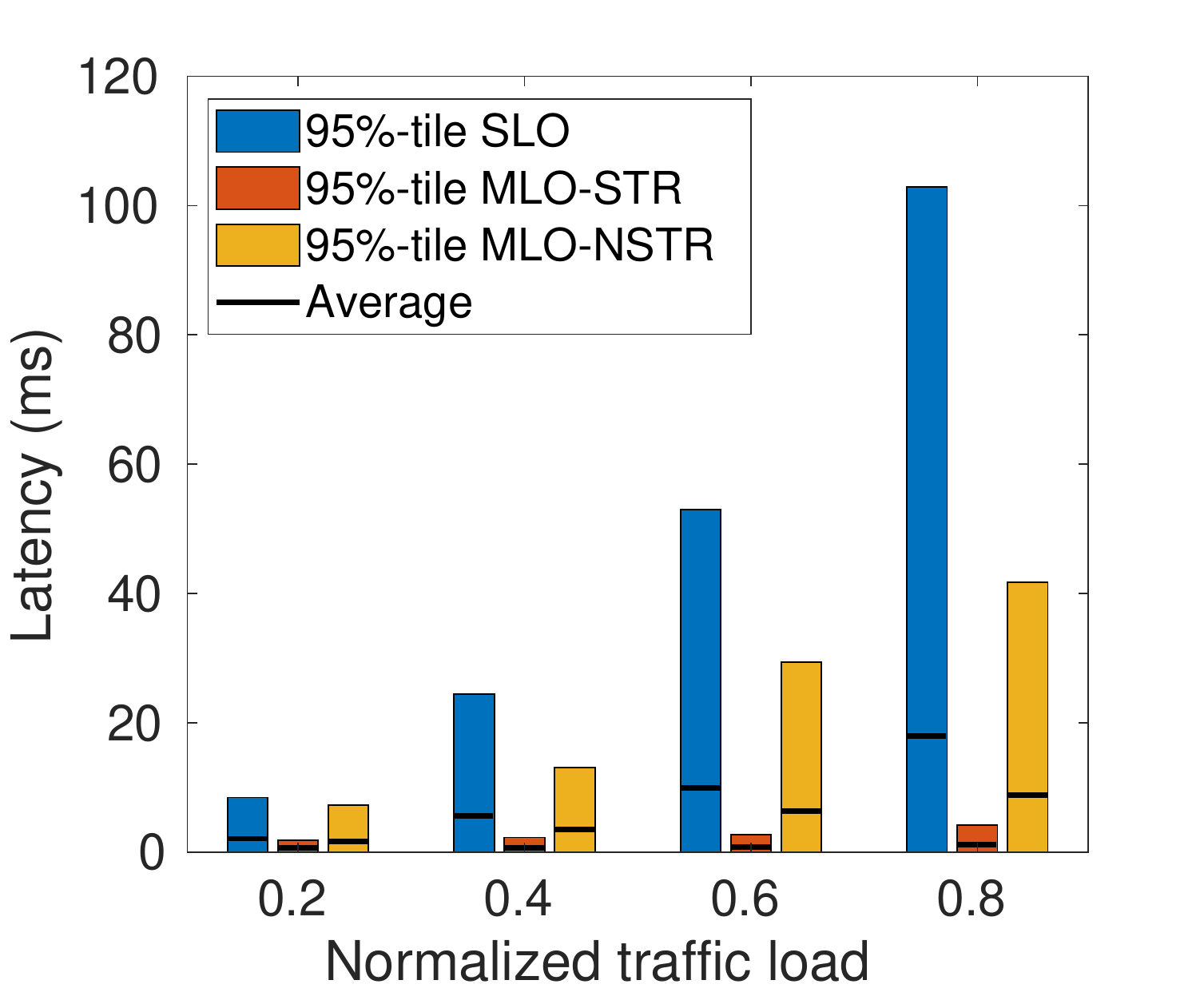}
    \caption{ 40\% occupancy on both links}
    \label{sym4}
\end{subfigure}
\begin{subfigure}[b]{0.325\textwidth}
    \includegraphics[width = \textwidth]{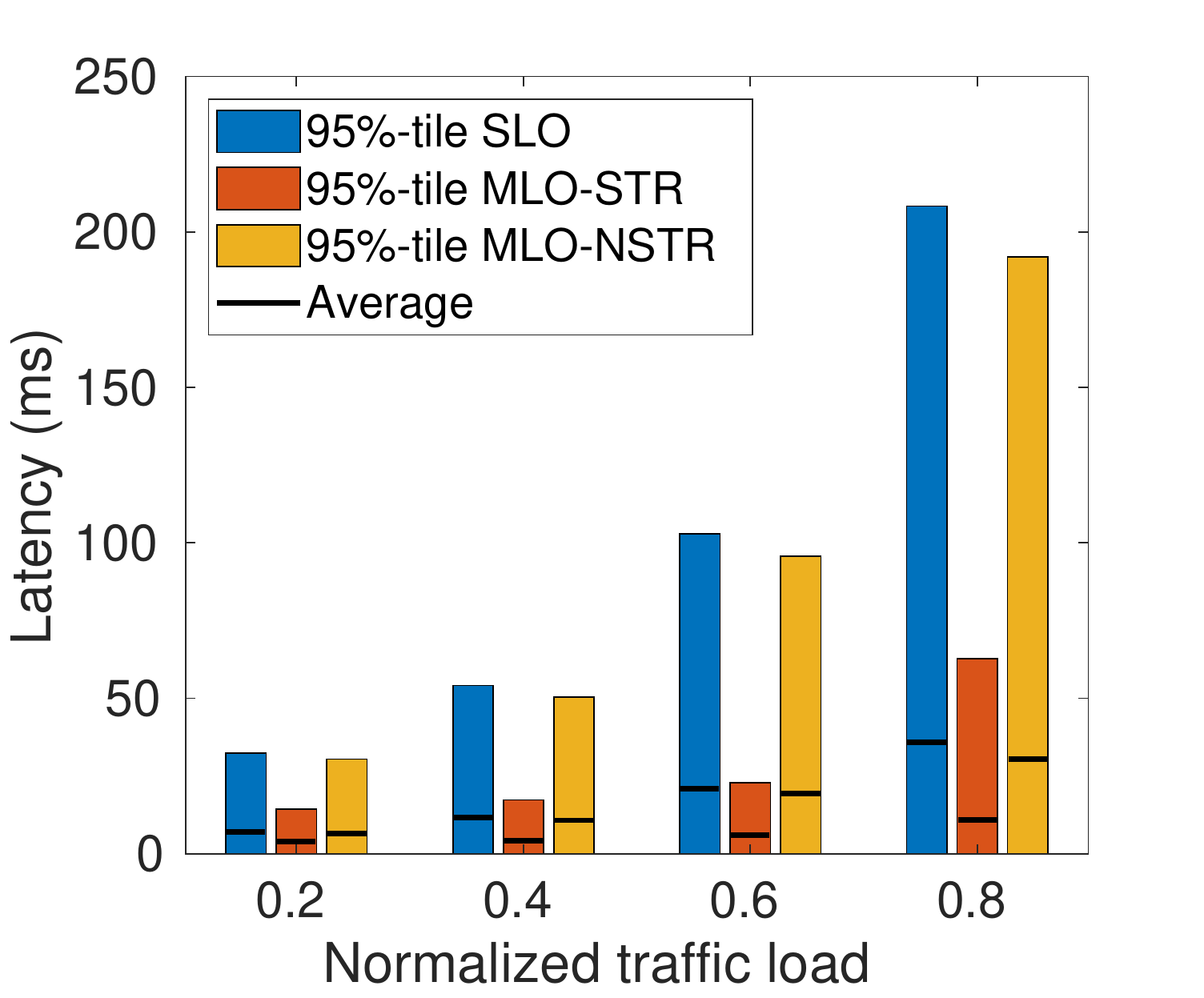}
    \caption{70\% occupancy on both links}
    \label{sym7}
\end{subfigure}
\caption{Latency for symmetrically occupied links vs.  normalized traffic load.}
\label{syms1}
\end{figure*}

We begin by studying the case of \emph{symmetric} link occupancy, in which both MLO links have channels with similar occupancy levels. In particular, we study epochs when both links have  occupancy of about 10\%, 40\%, and 70\%, which we denote as  symmetric low, medium, and high occupancy.
For SLO, the average throughput with fully backlogged traffic on the single link is 37~Mbps, 22~Mbps, and 6.8~Mbps, respectively. 
For MLO, we study the delay performance in the three spectrum occupancy cases as the traffic load increases. We consider four traffic loads of \{0.2, 0.4, 0.6, 0.8\} times the SLO throughput, so that all access modes operate in a non-saturated regime, hence allowing to study their delay in comparable conditions.

Figure~\ref{syms1} shows the average and 95th percentile delay for all channel access modes and the different link occupancies. First, observe that when both links have 10\% occupancy (Figure \ref{sym1}), MLO has strikingly improved latency scaling with increasing traffic load compared to SLO. For example, at 20\% traffic load, MLO-STR and MLO-NSTR offer a modest decrease in average delay compared to SLO of 17\% and 9\% respectively. In contrast, when the traffic load is 80\%, MLO-STR and MLO-NSTR reduce the average delay by 69\% and 62\%. This scaling is even more pronounced when analyzing the 95th percentile of delay, in which MLO  achieves up to a 78\% delay reduction. Thus, for both average and 95th percentile delay, the benefits of MLO are increasingly pronounced under higher traffic load. Indeed, in this case there are often multiple packets in the buffer such that both interfaces can be used. Moreover, with a relatively low link occupancy of 10\%, both links are often available.  

Next,  consider the case that both links have symmetrical medium (40\%) occupancy (Figure \ref{sym4}) and note the change in y-axis scale. Here,  SLO's \emph{average} delay increases by nearly an order of magnitude with increasing traffic (i.e., from 2 to 18~ms), whereas the \nnp delay increases much more rapidly, exceeding 100~ms. In contrast, MLO-STR can yield a striking order of magnitude reduction in \nnp delay compared to SLO. The reason is that with access to either or both links, MLO-STR realizes delay benefits unless both links are occupied. 

Unfortunately, unlike MLO-STR, the benefits of MLO-NSTR over SLO are limited and mostly confined to how the average delay scales with traffic load. Indeed, MLO-NSTR can only gain access to the secondary link if the primary link is also idle, implying that the average delay is guaranteed to be lower than the average delay under SLO. However, the \nnp delays are triggered by long periods of occupancy of the primary link, thus making any availability of the secondary link during this time irrelevant. As a result, the \nnp delay under MLO-NSTR rapidly grows as the normalized traffic load increases. 

Lastly, when both links have high (70\%) occupancy (Figure \ref{sym7}), MLO-STR again has the most favorable \nnp delay scaling with traffic load, providing substantial reductions as compared to both SLO and MLO-NSTR. Nonetheless, at such high link occupancies, even MLO-STR has difficulty finding transmission opportunities on either link, so both mean and \nnp delays are increasing. Additionally, MLO-NSTR provides negligible benefits in both average and \nnp delay compared to SLO.

\textit{Findings:} When both links have symmetric medium to high occupancy, MLO-NSTR fails to provide significant \nnp delay benefits compared to SLO. The key reason is that MLO-NSTR is only able to realize a benefit compared to SLO if both links are simultaneously unoccupied, an increasingly unlikely occurrence in this scenario. Fortunately,  MLO-STR yields significant \nnp latency benefits (compared to both SLO and MLO-NSTR) even in the challenging regime of increasing occupancies and traffic. This is because MLO-STR can reduce  packet waiting time even when it cannot simultaneously utilize both available links. 


\subsection{Asymmetrically Occupied links}\label{wcd}

\begin{figure*}[ht]
\centering
\begin{subfigure}[b]{0.325\textwidth}
    \includegraphics[width = \textwidth]{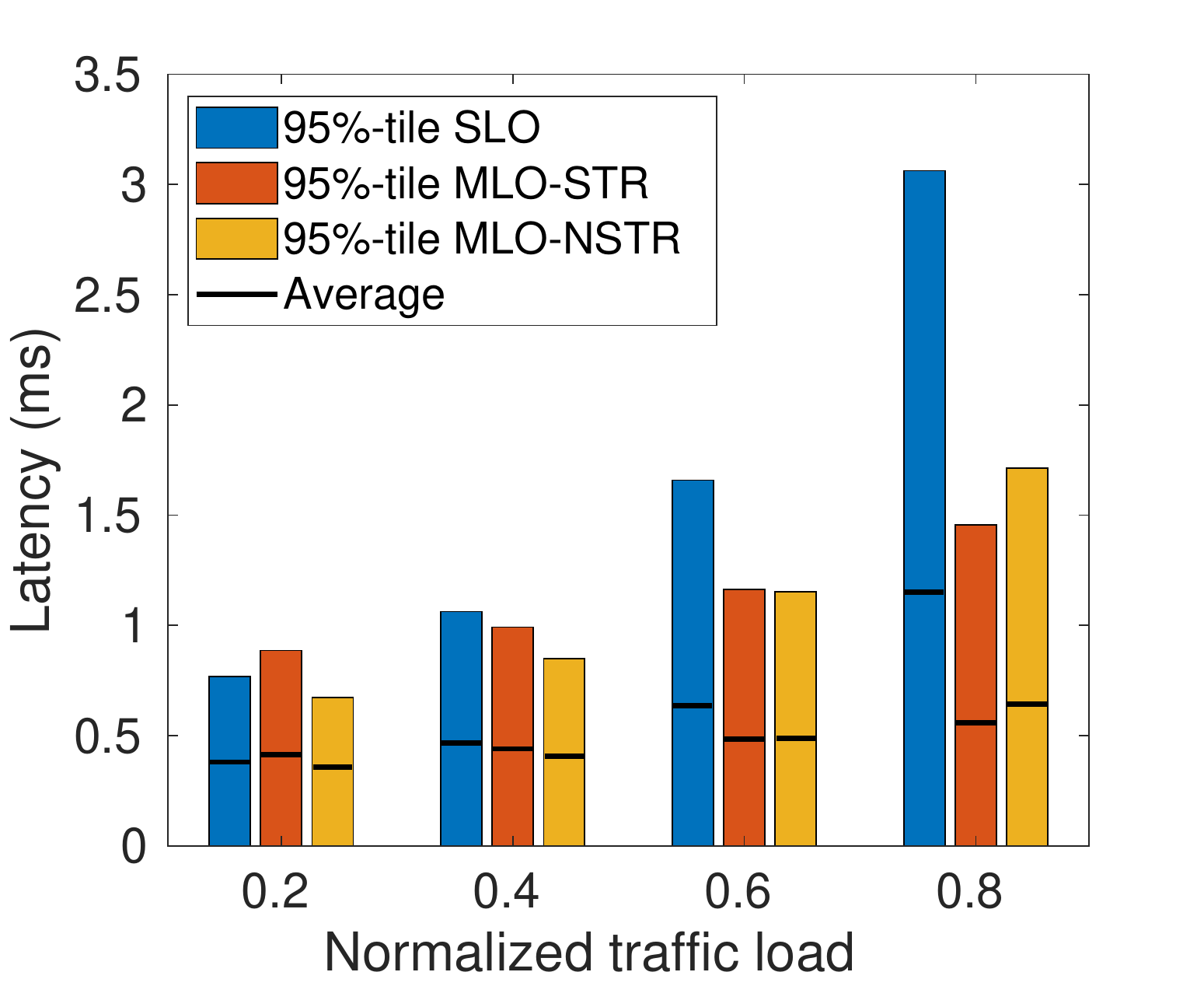}
    \caption{\{10\%, 40\%\} occupancy.}
    \label{delay1n}
\end{subfigure}
\begin{subfigure}[b]{0.325\textwidth}
    \includegraphics[width = \textwidth]{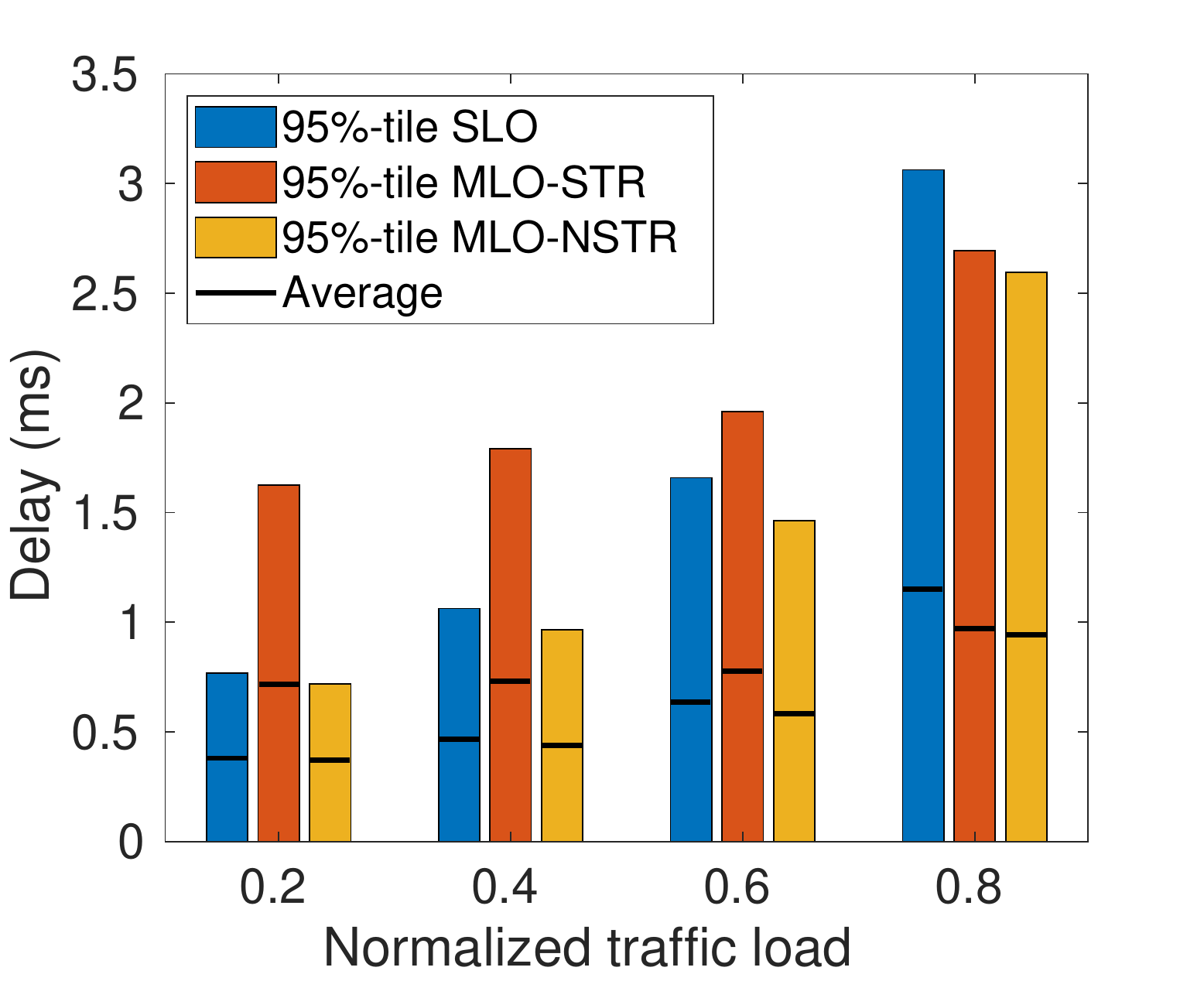}
    \caption{\{10\%, 70\%\} occupancy.}
    \label{delay3n}
\end{subfigure}
\begin{subfigure}[b]{0.325\textwidth}
    \includegraphics[width = \textwidth]{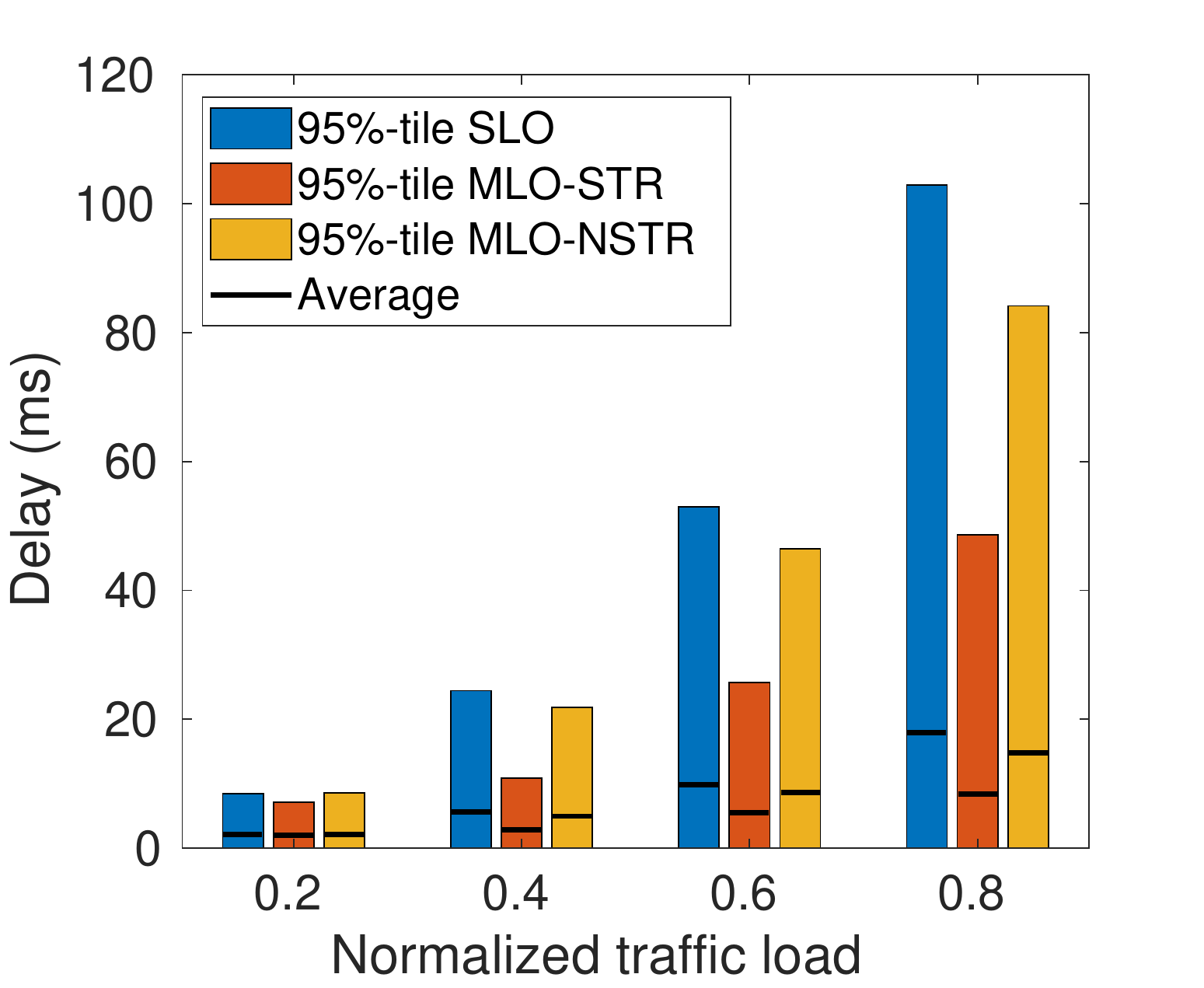}
    \caption{\{40\%, 70\%\} occupancy.}
    \label{delay5n}
\end{subfigure}
\caption{Latency for asymmetrically occupied links (indicated as \{primary, secondary\}) vs.  normalized traffic load.}
\label{delsn}
\end{figure*}

We continue by employing the same normalized traffic loads as in the previous section, but consider epochs of asymmetric  link occupancy. Between the two links, we present the case that the primary is the less occupied one, and gauge to what extent each MLO mode can exploit an extra (albeit busier) link to reduce the delay. We note that the opposite case---i.e., when the secondary link is on average less busy than the primary---favors both MLO modes, since SLO would always incur a high delay and both MLO modes would take advantage of a more idle secondary link.

Figure~\ref{delay1n} depicts the case of a low (10\%) primary and medium (40\%) secondary link occupancy. As expected, MLO-NSTR always offers  lower delay than SLO, with the highest benefits occurring under higher traffic loads. However, MLO-STR surprisingly incurs a higher average and \nnp delay than SLO for the lowest normalized traffic load of 0.2. Indeed, MLO-STR starts contention by initializing the backoff counter as soon as a link is detected to be idle. Unfortunately, such a link may be occupied before the backoff timer expires, thus pausing the backoff counter. If the backoff is paused too often (or for long intervals), the packet could incur even higher delays than it would have if the other link---initially busy---had been selected.

In Figure~\ref{delay3n}, this effect is exacerbated due to the even higher occupancy of the secondary link, as selecting an idle secondary link incurs the risk of the latter being occupied before the backoff counter expires. When this occurs, the \nnp delay can be twice as high as that with SLO, albeit still confined to below 10~ms. Nonetheless, MLO-STR average and \nnp delays grow at a lower rate than those of SLO as the traffic load increases. Indeed, MLO-STR can still take advantage of a secondary link (even when highly occupied) to reduce congestion and curb the latency when it is caused not only by the link occupancy patterns but also by the amount of traffic.

Finally, Figure \ref{delay5n} depicts  primary and secondary link occupancy of 40\% and 70\%, respectively (note the different y-axis scale due to higher load). Similar to the symmetric cases of Figures~\ref{sym4} and \ref{sym7}, MLO-STR scales well with  increasing traffic load, keeping the average delay below that of SLO and decreasing the 95th percentile by up to a half. Compared to Figure~\ref{delay3n}, because the primary link occupancy has increased,  SLO delay increases more sharply with traffic load, whereas  MLO-STR achieves the lowest delays for the same reasons as in the symmetric case.

\textit{Findings:} Asymmetric primary vs. secondary link occupancy radically transforms MLO performance. For MLO-STR,  despite its uniformly superior performance in symmetrically occupied links, a secondary link that is much busier than the primary can lead to even higher delays than using SLO. This is owed to packets being suboptimally assigned to an interface before carrying out the backoff, with the latter likely to be interrupted on the busier link. This effect is exacerbated when the difference between link occupancies increases.


\section{MLO-STR Latency Decomposition and Reduction}\label{decomposition}

Given MLO-STR's superior throughput performance and delay performance under symmetrically occupied links, we proceed to study the origins of the surprisingly worse delay than SLO under asymmetric links. Moreover, we demonstrate how to overcome this limitation with a modification to MLO-STR.

\subsection{Access and Queueing Delay}

\begin{figure*}
    \centering
    \begin{subfigure}[b]{0.45\textwidth}
        \includegraphics[width=\textwidth]{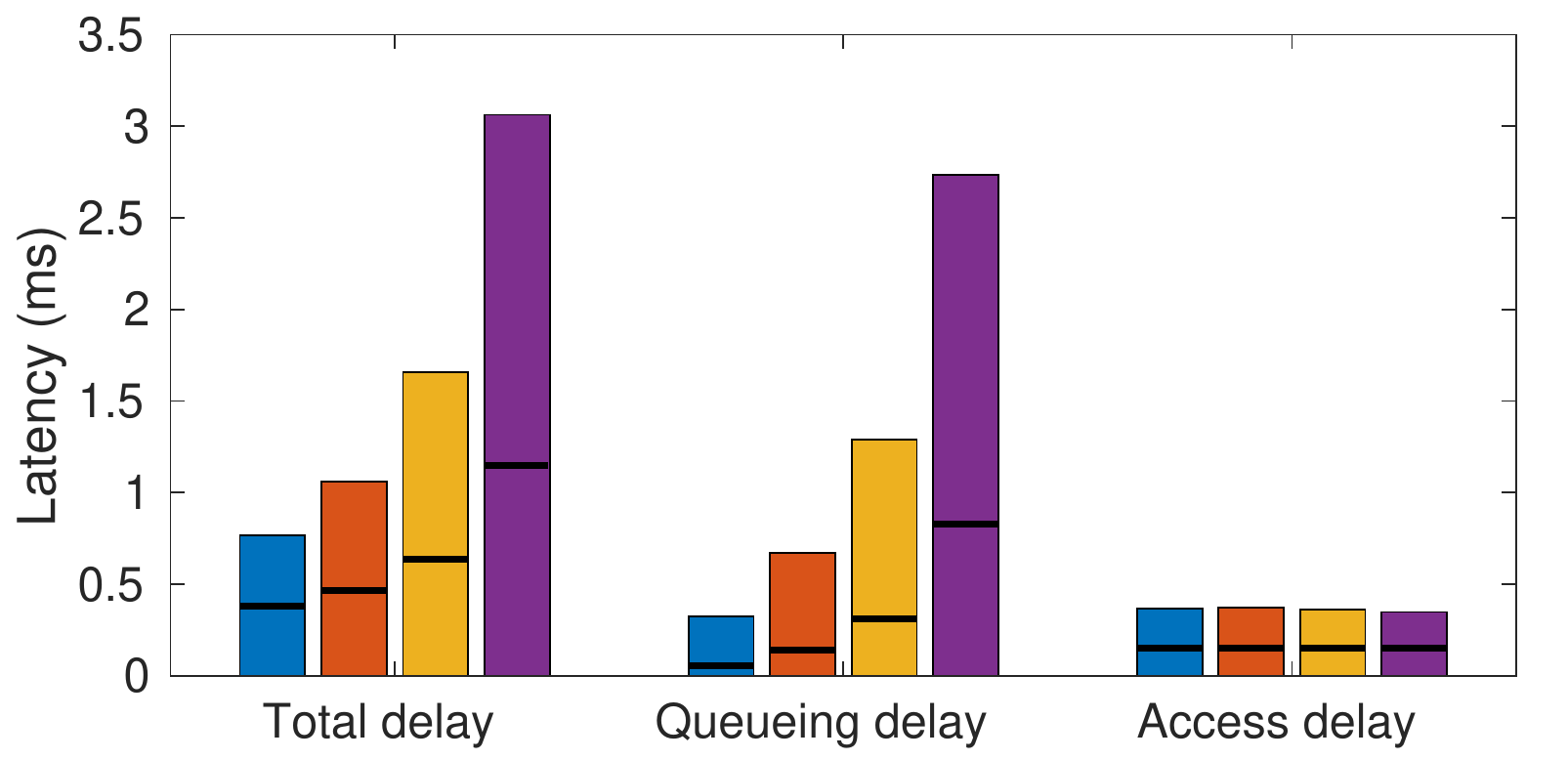}
        \caption{SLO latency for occupancy \{10\%, 10\%\}.}
        \label{1010}
    \end{subfigure}
    \begin{subfigure}[b]{0.45\textwidth}
        \includegraphics[width=\textwidth]{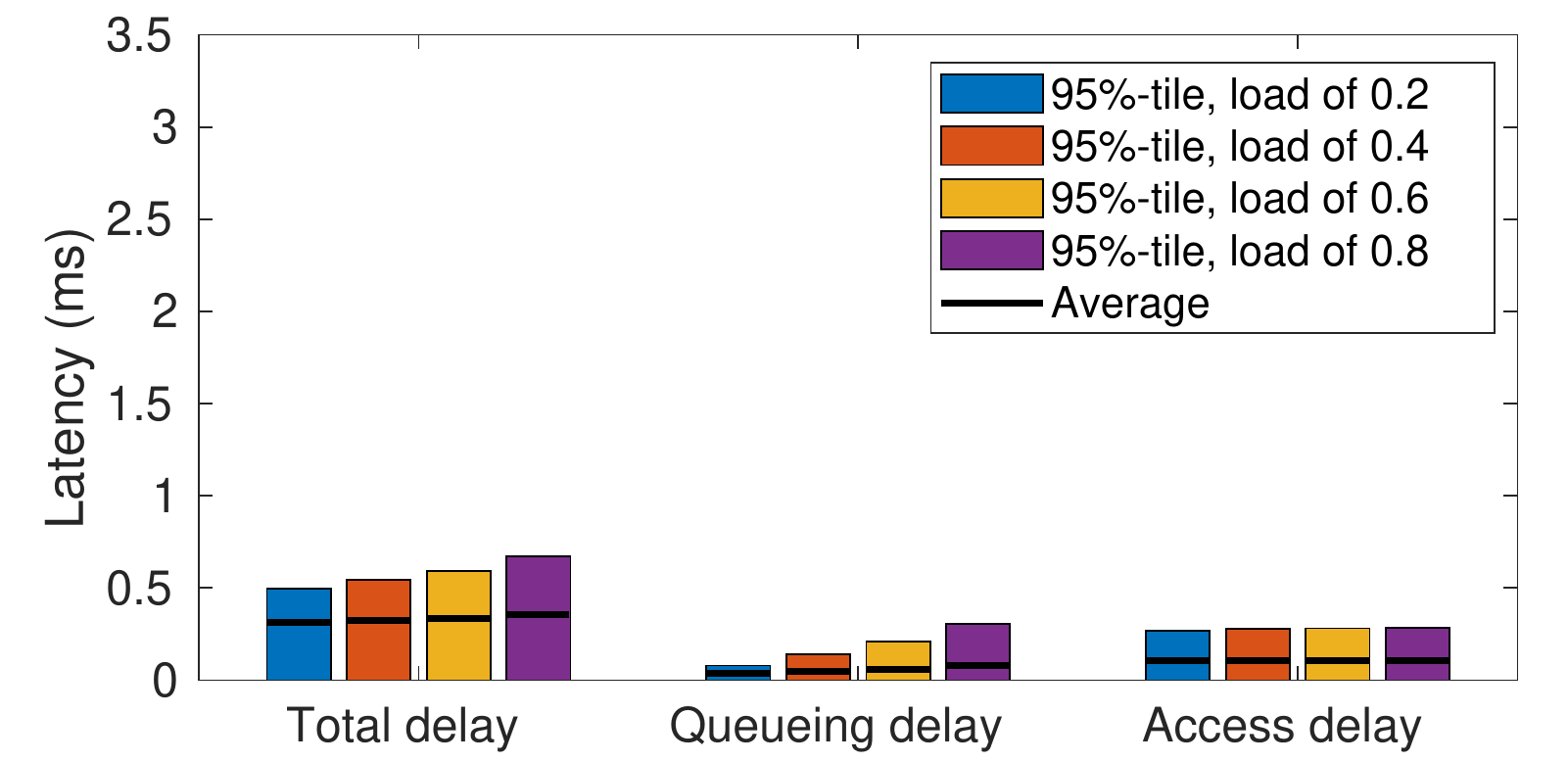}
        \caption{MLO-STR latency for occupancy \{10\%, 10\%\}.} 
        \label{1010-2}
    \end{subfigure}
     \begin{subfigure}[b]{0.45\textwidth}
        \includegraphics[width=\textwidth]{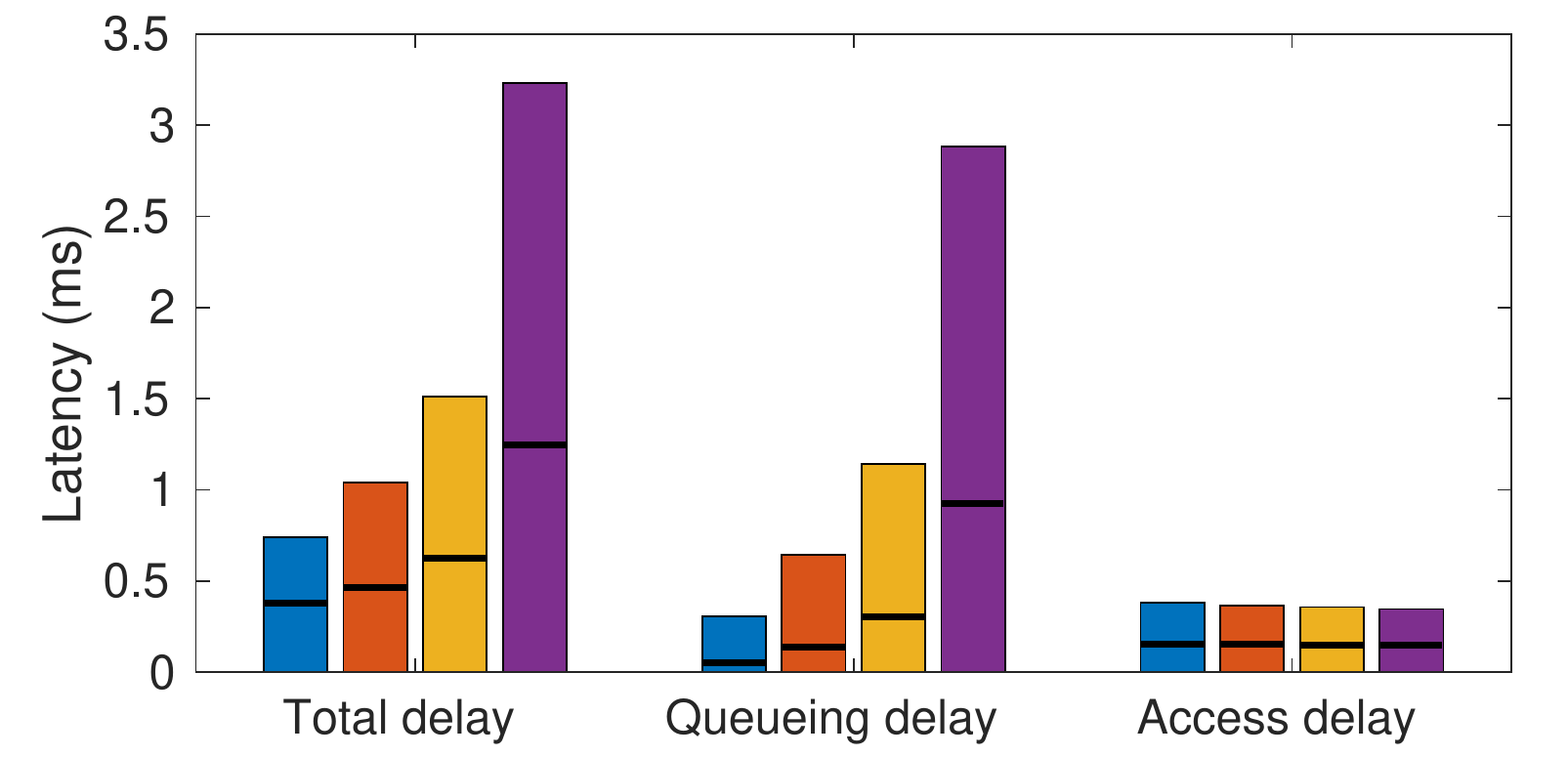}
        \caption{SLO latency for occupancy \{10\%, 70\%\}.} 
        \label{1070}
    \end{subfigure}
    \begin{subfigure}[b]{0.45\textwidth}
        \includegraphics[width=\textwidth]{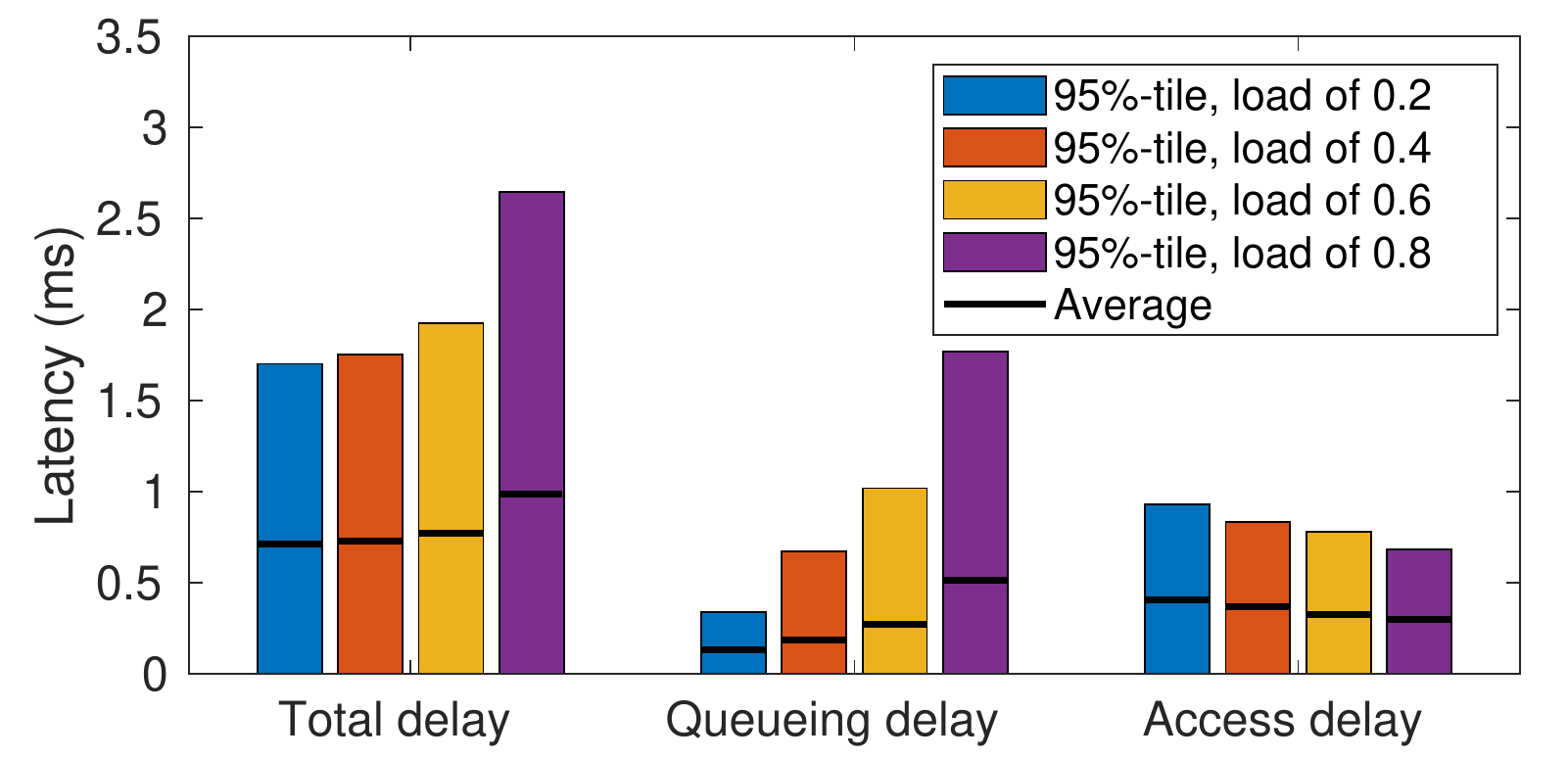}
        \caption{MLO-STR latency for occupancy \{10\%, 70\%\}.} 
        \label{1070-2}
    \end{subfigure}
    \caption{Access and queueing delay for SLO and MLO-STR.}
    \label{fig:slostr1066}
\end{figure*}

To study MLO-STR's latency performance, we decompose the total MAC delay into \emph{(i)} the \emph{queueing delay}, referring to the time elapsed since a packet arrives to the AP until it is allocated to an interface and \emph{(ii)} the \emph{access delay}, that spans the time between a packet being assigned to an interface and the beginning of its transmission, i.e., the contention time.

Figures \ref{1010} and \ref{1010-2} compare the delay under SLO and MLO-STR when using links with \emph{symmetric occupancy} of $10\%$. Here, access delay remains largely unchanged between the two schemes across all traffic loads. Yet, the same does not hold for queueing delay, which remains low under MLO-STR for increasing load, whereas it rapidly grows for SLO, exceeding the access delay. Specifically, while for SLO the 95th percentile queueing delay can reach up to $9\times$ the access delay, MLO-STR curbs the waiting time and thus the total delay.

Figures~\ref{1070} and \ref{1070-2} keep the primary link occupancy to 10\% but now consider an \emph{asymmetric} secondary occupancy of 70\%. Although MLO-STR reduces the queueing delay with respect to SLO by availing of two interfaces, its access delay is higher than SLO, owing to a subset of packets being transmitted through the secondary link, which incurs a higher average occupancy. This latter effect may outweigh the former, resulting in higher total delay under MLO-STR than under SLO.
It can also be observed that MLO-STR access delay decreases as    traffic load increases. This is a side effect of the packet allocation process, explainable as follows. When a packet is assigned to the busier---but occasionally idle---secondary interface, it forces multiple subsequent packets through the primary interface as the only option. The latter creates an inherent trade-off in that for every packet assigned to the busier interface, multiple others are assigned to the interface having lower occupancy. As the traffic load increases, this phenomenon becomes more pronounced, leading to more packets being assigned to the lesser-occupied primary link, thus reducing the average and the \nnp access delay.

\textit{Findings:} MLO-STR significantly reduces the time that packets spend waiting in the queue when compared to SLO, for both symmetrically and asymmetrically occupied links. However, while the access delay is similar when the two links are symmetrically occupied, it can increase for MLO-STR when the secondary link is busier than the primary, sometimes outweighing the benefits of a reduced waiting time. These events of high access delay are the result of assigning the packet to an interface before the backoff starts.  Indeed, a packet may be assigned to an interface just before a long link occupancy phase, pausing the backoff counter repeatedly and significantly delaying the packet transmission.


\subsection{MLO-STR with Deferred Decision}

We have shown that---and explained why---MLO-STR can lead to even higher delays than  SLO in the case of links with different occupancies. 
To better understand the design space of MLO channel access, we  define a minor variation---denoted \emph{MLO-STR with Deferred Decision} (MLO-STR+)---which   overcomes this limitation of MLO-STR by deferring the decision about which link to use until the end of the backoff countdown. 

\begin{figure}[ht!!!]
    \centering
    \includegraphics[width=\columnwidth]{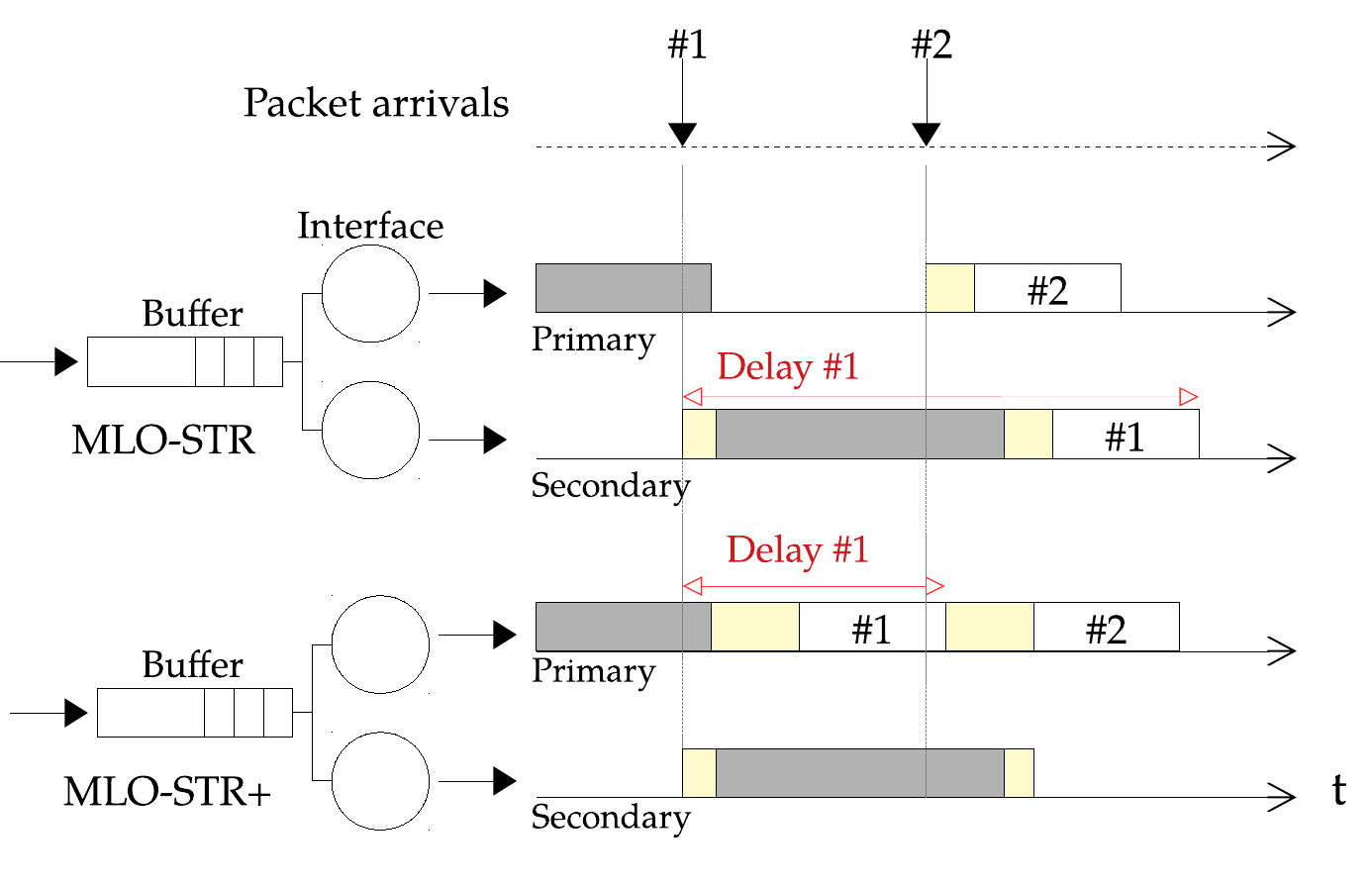}
    \caption{Illustration of MLO-STR vs MLO-STR+. MLO-STR+ reduces the delay incurred by packet \#1 at the expense of that incurred by packet \#2.}
    \label{Fig:MLO_channelaccess_plus}
\end{figure}

The main features of MLO-STR+ are \emph{(i)} running as many backoff instances as radio interfaces while there are packets waiting for transmission, and \emph{(ii)} allocating the first packet waiting to the interface with the backoff counter that expires first. This differs from MLO-STR, in which we allocate packets once links are idle, before running the backoff. Implementing MLO-STR+  requires only minor changes to the Wi-Fi MLO state machine: the ability to control when an interface can initiate, pause, and complete the backoff countdown without actually being allocated a packet. A comparison between MLO-STR and MLO-STR+ operations is illustrated in Figure~\ref{Fig:MLO_channelaccess_plus}, where it can be observed that the delay experienced by the first packet (the one exhibiting the worst-case delay in MLO-STR), is notably reduced with MLO-STR+ albeit at the expense of increasing the delay of the second packet.

\begin{figure*}[ht]
\centering
\begin{subfigure}[b]{0.325\textwidth}
    \includegraphics[width = \textwidth]{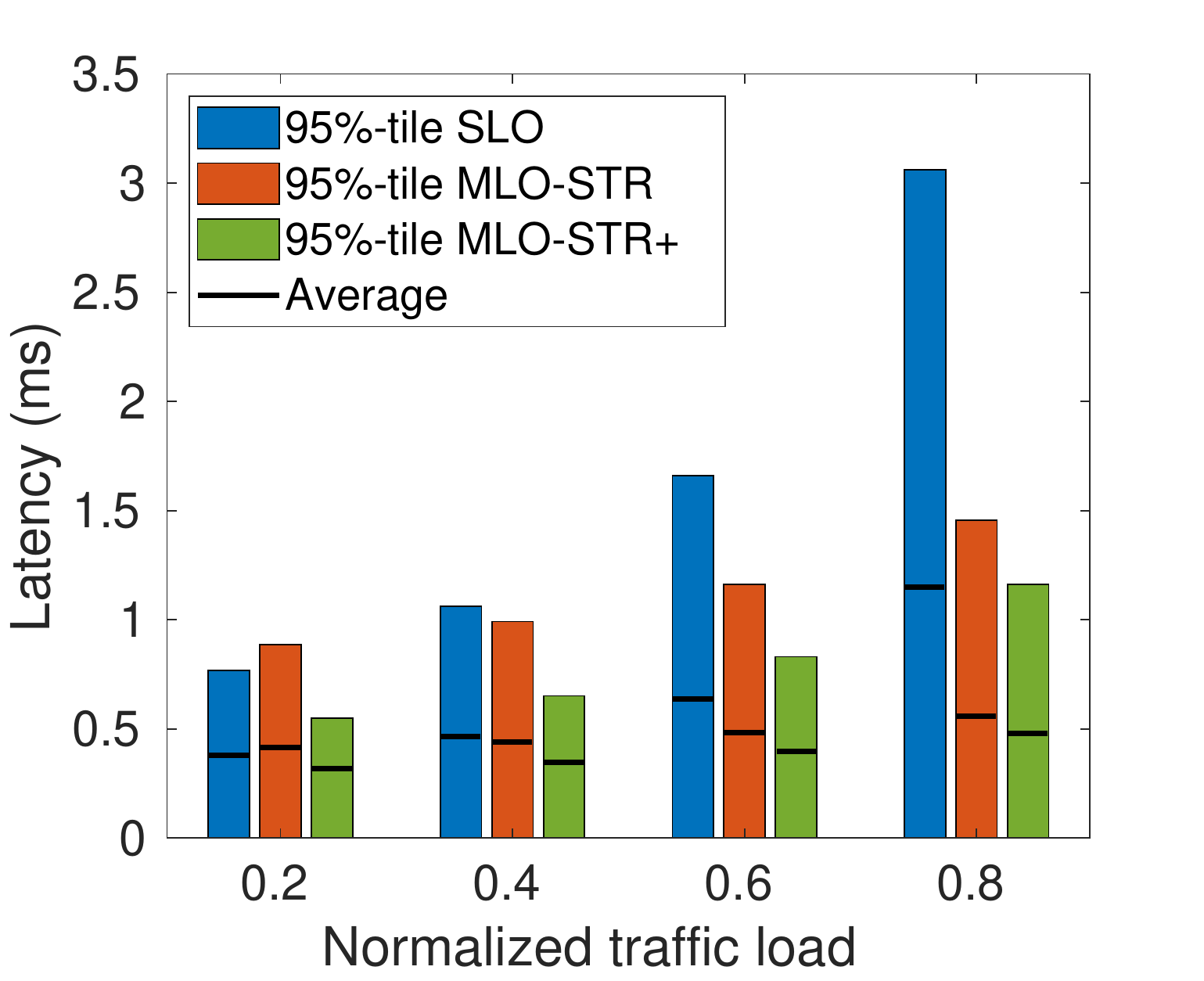}
    \caption{Link occupancy of \{10\%, 40\%\}.}
    \label{delay+}
\end{subfigure}
\begin{subfigure}[b]{0.325\textwidth}
    \includegraphics[width = \textwidth]{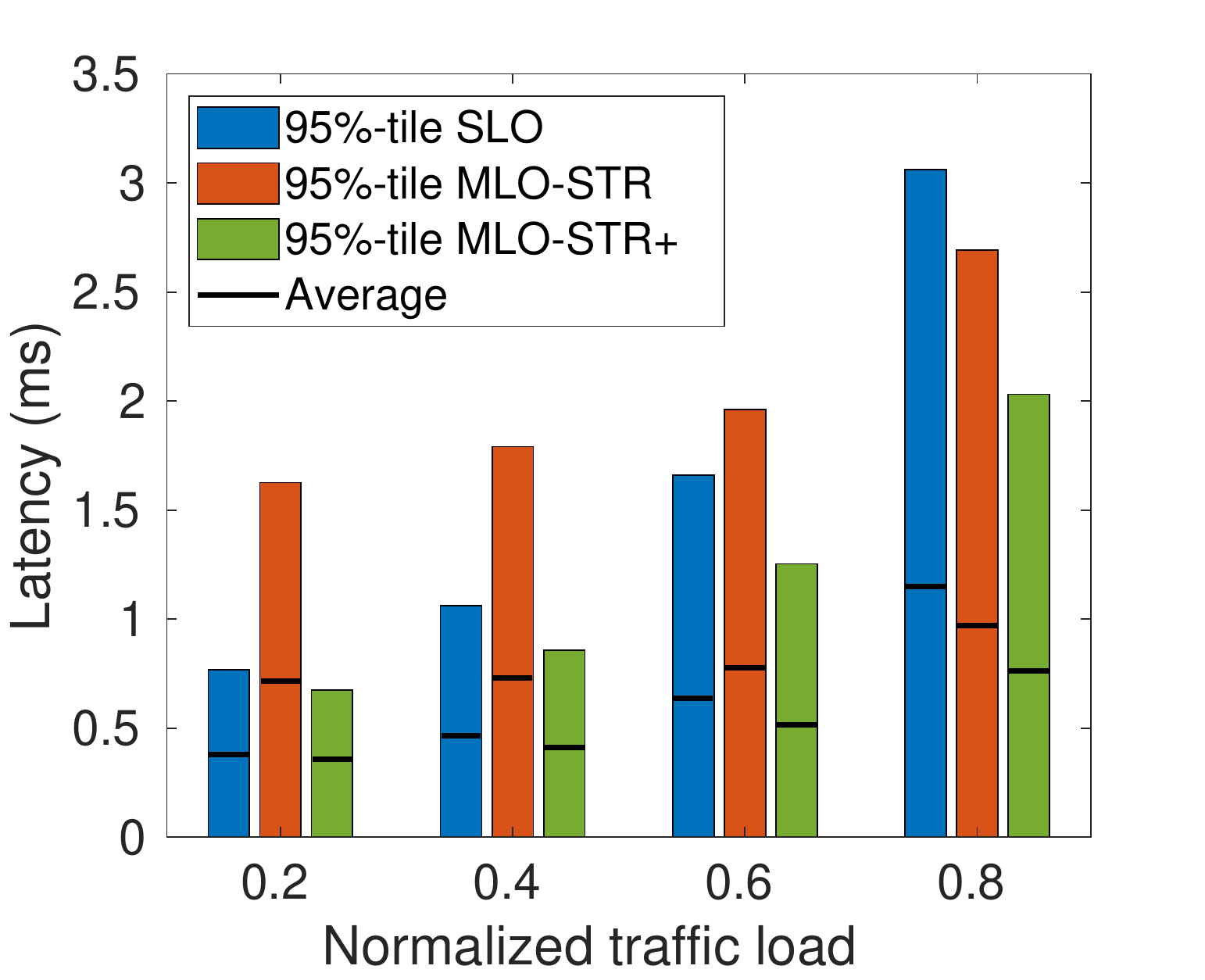}
    \caption{Link occupancy of \{10\%, 70\%\}.}
    \label{nine}
\end{subfigure}
\begin{subfigure}[b]{0.325\textwidth}
    \includegraphics[width = \textwidth]{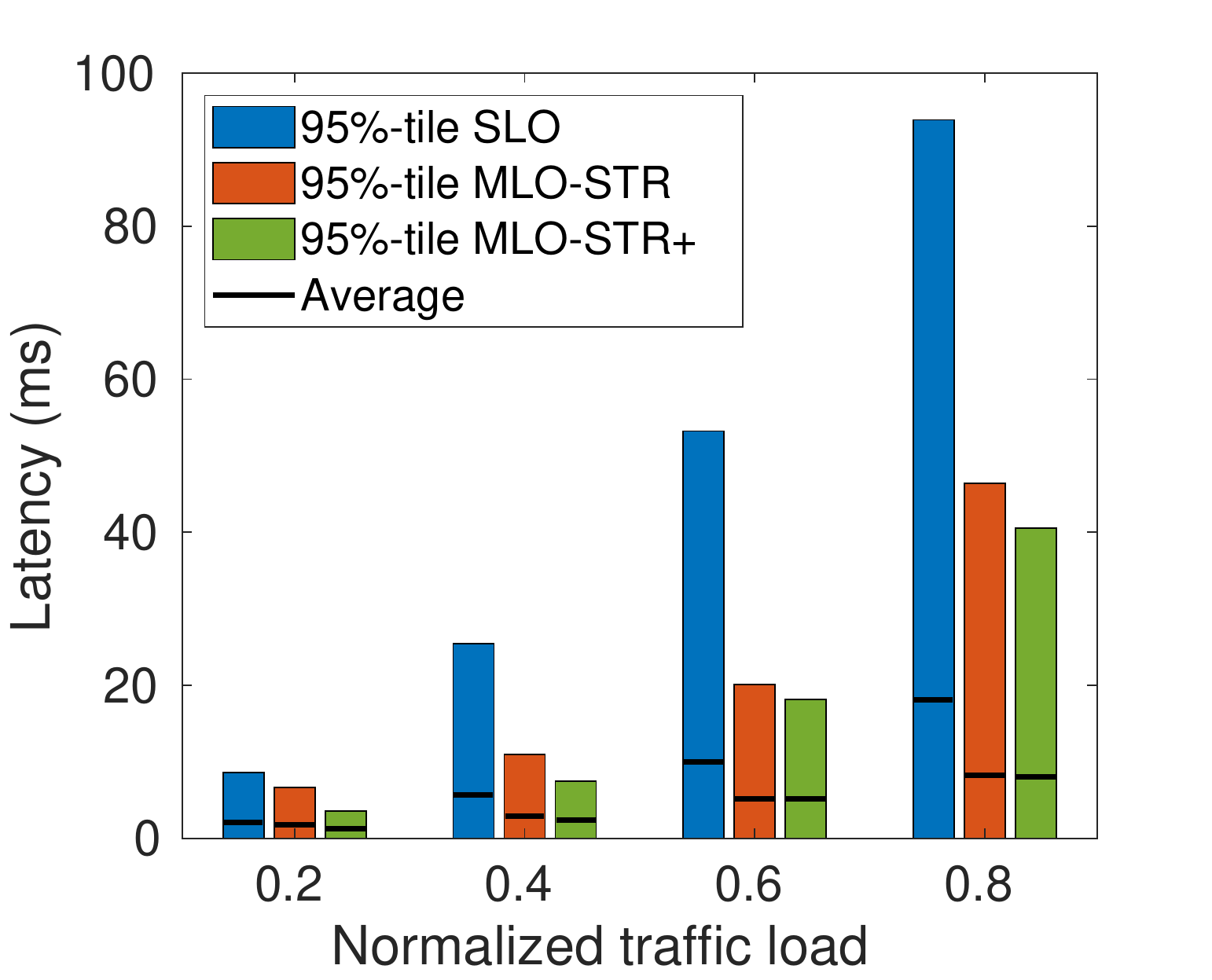}
    \caption{Link occupancy of \{40\%, 70\%\}.}
    \label{delay3+}
\end{subfigure}
\caption{Delay for asymmetric links vs. variable normalized traffic load under SLO, MLO-STR, and MLO-STR+.}
\label{delsn+}
\end{figure*}

\begin{figure}[ht]
    \centering
        \includegraphics[width=0.45\textwidth]{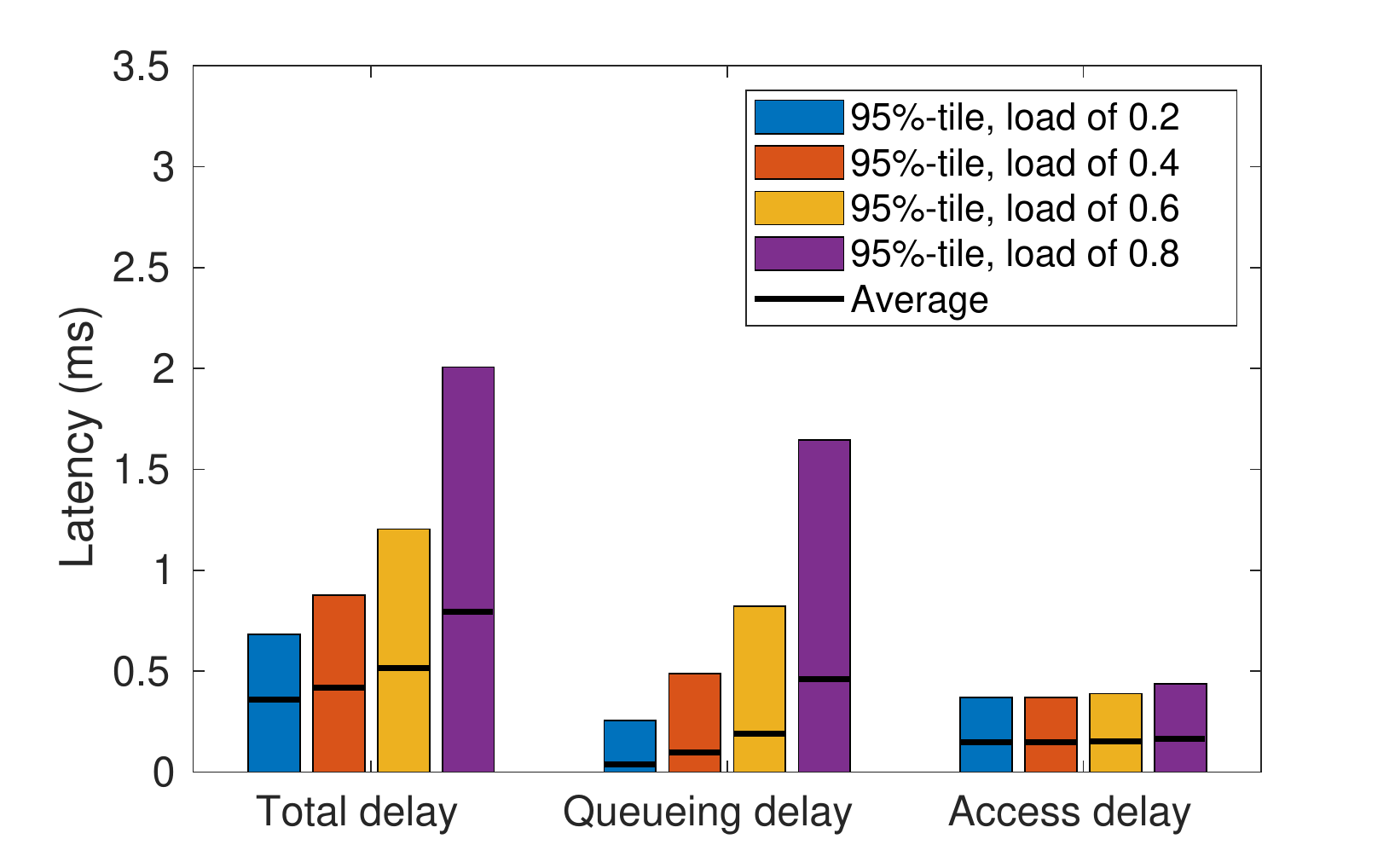}
        \caption{Delay for MLO-STR+ with \{10\%, 70\%\} occupancy.}
        \label{10702}
    \label{fig:my_label2}
\end{figure}

Figure~\ref{delsn+} shows the resulting average and 95th percentile delay for MLO-STR+,    MLO-STR, and SLO. Observe that MLO-STR+   consistently outperforms both MLO-STR and  SLO for both average and 95th percentile delay for all traffic loads and link occupancy combinations. Figure \ref{fig:my_label2} depicts the breakdown of the waiting and access delay of MLO-STR+, showing that MLO-STR+ reduces the average and worst-case access delays while maintaining similar queueing delays than those in Figure~\ref{1070}. Namely, the \nnp of both the access delay and the total delay are reduced by up to $60\%$ when employing MLO-STR+ instead of MLO-STR.

\textit{Findings:} MLO-STR+ improves  MLO-STR by delaying the link allocation of the packet at the head of the queue until one of the backoff counters expires, thereby leveraging up-to-date information on the link state, and  ultimately improving the link allocation  decision. The performance of MLO-STR+ also shows how MLO channel access can be improved by jointly controlling the operation of the multiple radio interfaces.

\subsection{Packet jitter}

Variance in the delay is an important metric to consider for next-generation networks. We have shown that MLO can be used to reduce network delay, and expect that it can also reduce the variability with which transmissions reach their target. 

\begin{figure}[h]
    \centering
    \includegraphics[width = 0.45\textwidth]{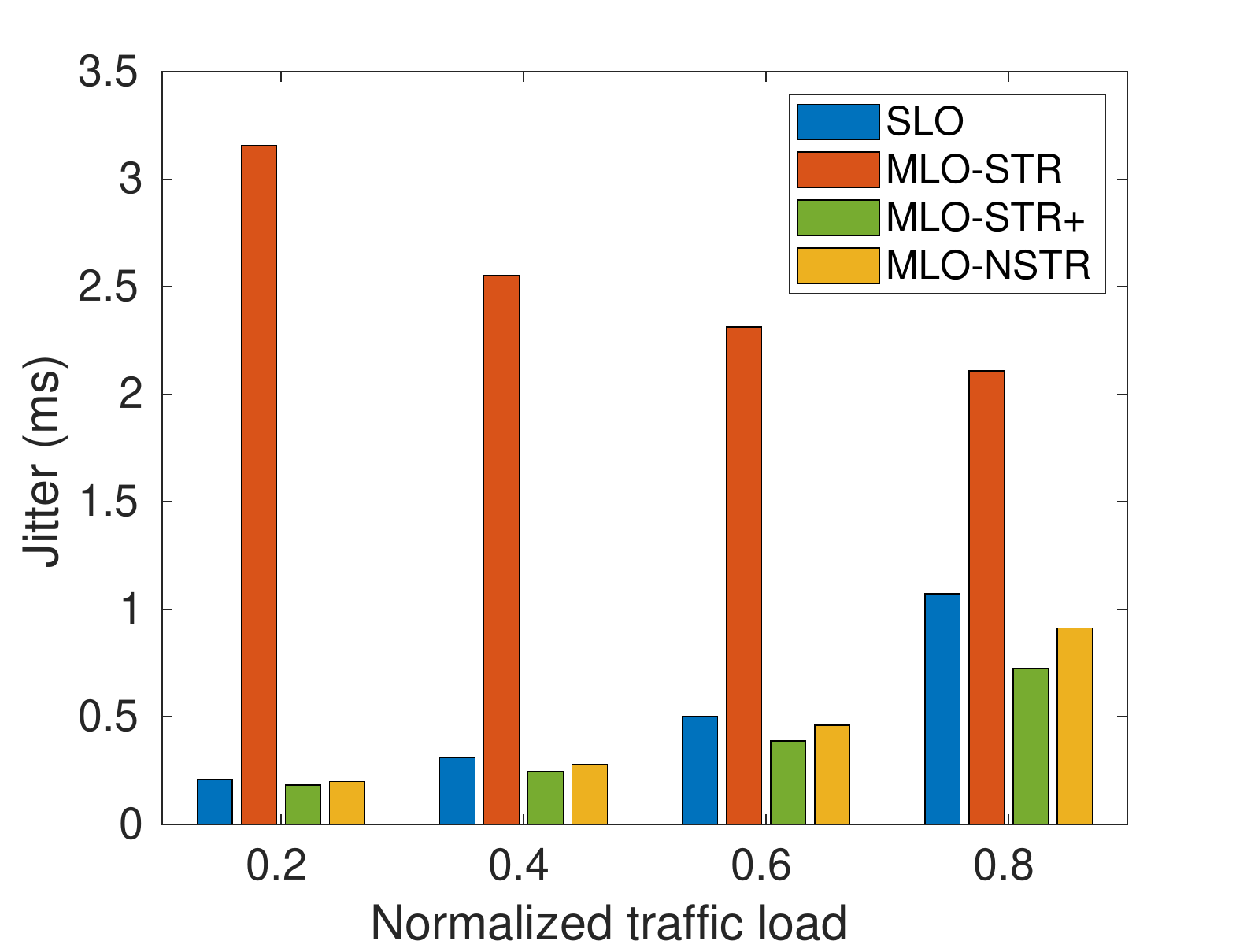}
    \caption{Packet jitter for \{10\%, 70\%\} occupancy.}
    \label{jitter}
\end{figure}

We analyse the packet jitter for the same cases shown in Figures  \ref{delay3n} and \ref{nine}, by studying the standard deviation of the delay. Figure \ref{jitter} shows the jitter for each access method. The jitter increases with traffic load for all access methods, except for MLO-STR, where decreases steadily. The reason for such a behavior in the case of MLO-STR is the high delays some packets suffer when transmitted through the secondary link, which at low traffic loads significantly differ from the low delay of the packets transmitted through the primary link. For a traffic load of 0.8, the only point in which SLO has a higher delay than MLO-STR (see Fig. \ref{nine}), MLO-STR still shows a jitter that is twice as high as the one for SLO. This high variability in MLO-STR delay further showcases its limitations in assigning packets to interfaces, and the negative effect of allocating packets to a busy secondary link. Similar as in the delay, MLO-STR+ is able to circumvent such a situation by deferring the decision about which link to use at the end of the backoff countdown.

\textit{Findings:} MLO-STR incurs a high variability in its delay due to the blind allocation of packets to high occupied links, having jitter an order of magnitude higher than any other method at worst. MLO-STR+ and MLO-NSTR keep their jitter below that of SLO, thus offering more consistency in their operation.


\section{Delay Experiments with Real Traffic}
\label{sec:GoogleStadia}

\begin{figure*}[ht]
\centering
\begin{subfigure}[b]{0.4\textwidth}
    \includegraphics[width = \textwidth]{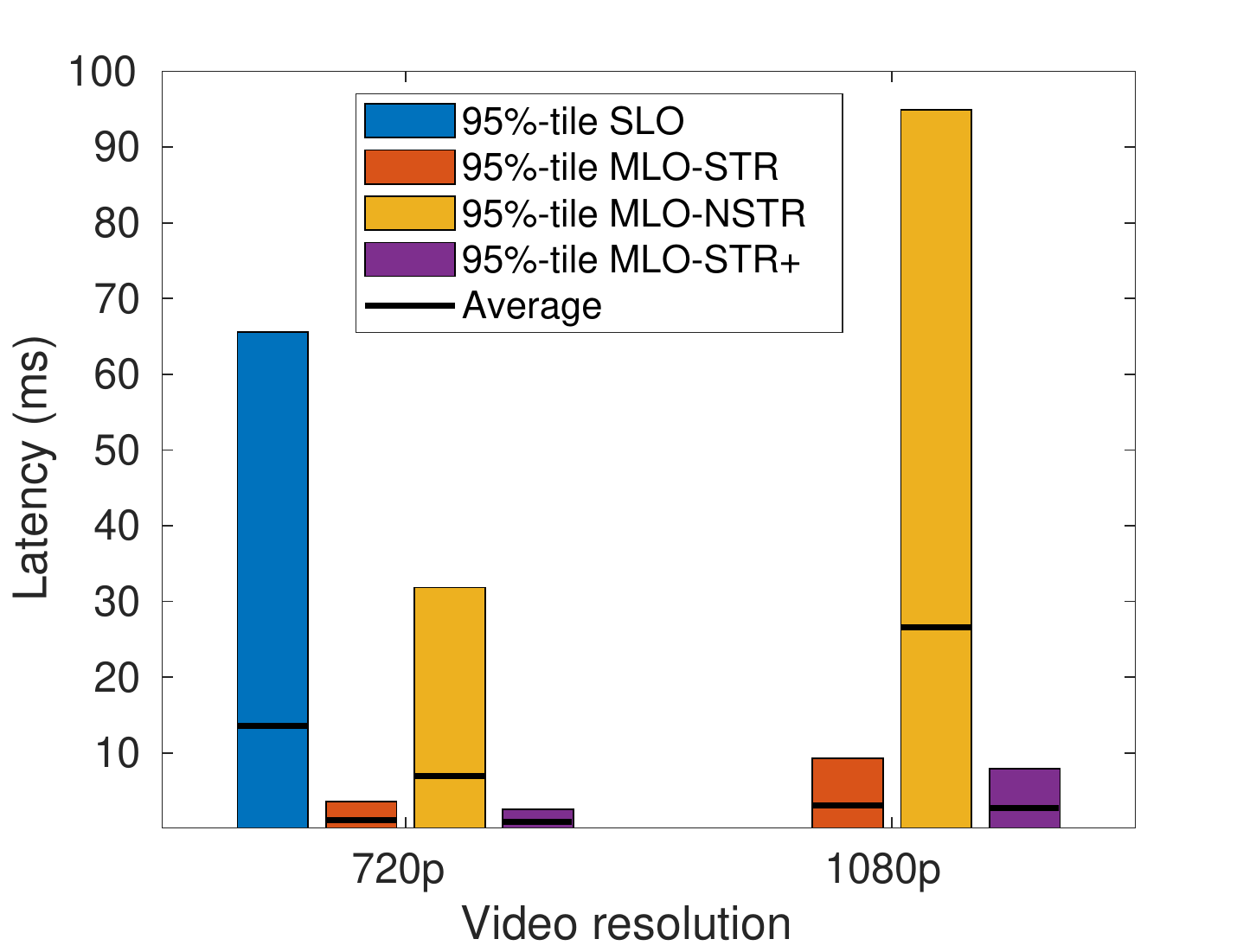}
    \caption{Primary 40\% and secondary 40\% occupancy}
    \label{st2}
\end{subfigure}
\begin{subfigure}[b]{0.4\textwidth}
    \includegraphics[width = \textwidth]{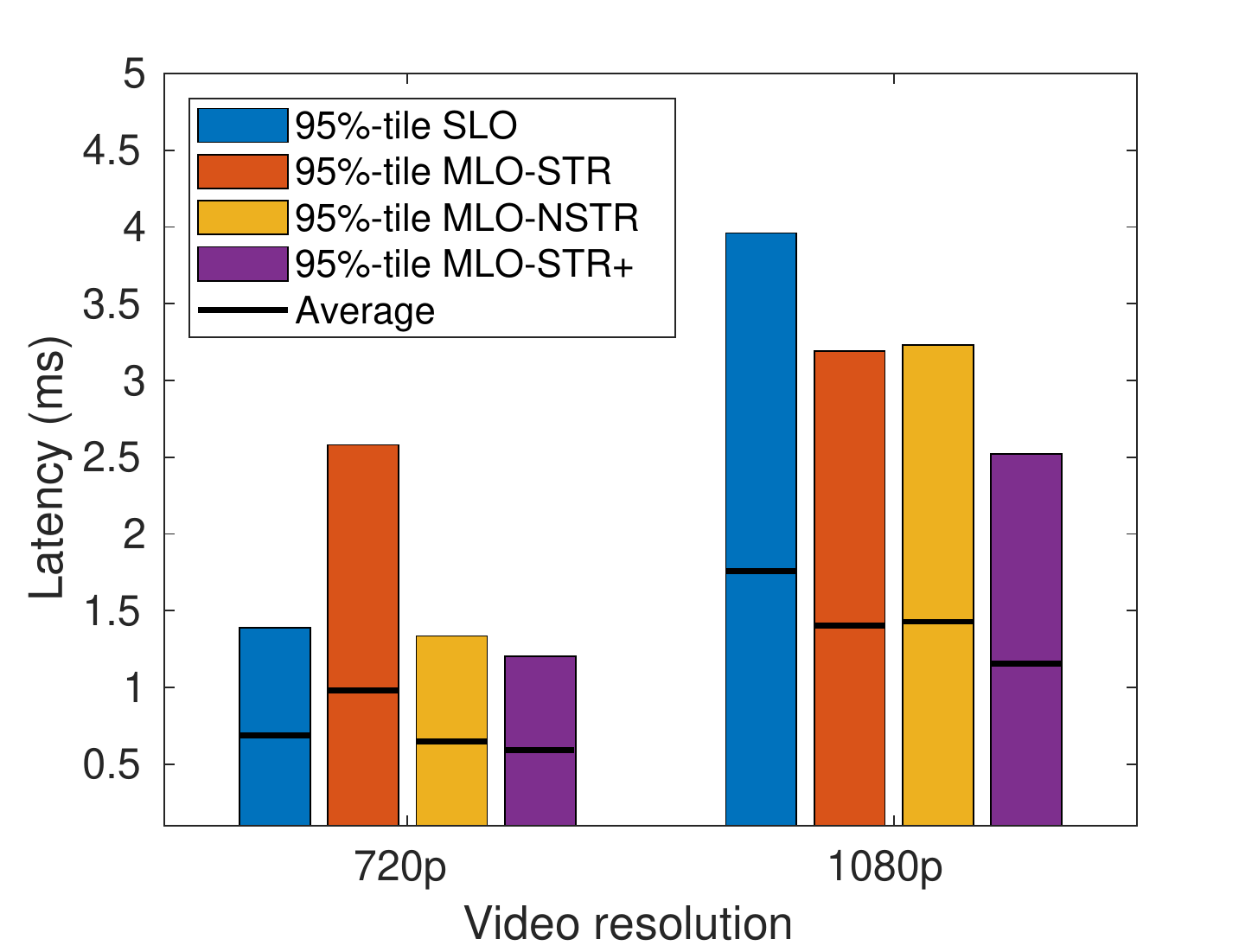}
    \caption{Primary 10\% and secondary 70\% occupancy}
    \label{st1}
\end{subfigure}
\caption{Delay vs. video resolution using Google Stadia traffic for (a) symmetric and (b) asymmetric link occupancy.}
\label{dst}
\end{figure*}

The experiments presented so far have used over-the-air channel traces coupled with artificial traffic generated by a Poisson Process with fixed packet size and variable rate.

Here, we consider actual application traffic traces in order to jointly account for real traffic and real channels. In particular, we employ Google Stadia \cite{carrascosa2022cloud}, a latency-sensitive cloud gaming application that streams videogames from Google's servers directly to a user's browser, and use it to generate packet size and arrival times.
We consider a single gamer and collect traces corresponding to two different screen resolutions, namely 720p and 1080p, leading to average traffic loads of $10.4$ Mbps and $22.1$ Mbps, respectively.
We use the same previous experimental methodology,  generating the traffic (packet arrival time and size) according to the values extracted from the captured Google Stadia's traces. The main difference between Poisson traffic  and Google Stadia's traffic is that, in the latter, packets arrive in periodic batches every 1/60 sec following the video frame rate, thus creating short congestion periods. More details about Stadia's traffic properties can be found in \cite{carrascosa2022cloud,  bellalta2020low, bellalta2022analysis} where they have been also employed to generate traffic as done in this paper.

Figure~\ref{st2} shows the delay performance for the same scenario as in Figure~\ref{sym4}, i.e., the one with symmetric link occupancy of $40\%$. {To facilitate a comparison between Figures~\ref{st2} and \ref{sym4}, we note that a normalized traffic load of $0.6$ and $0.8$ from Figure \ref{sym4} is the closest to the traffic load of Stadia's 720p and 1080p resolutions respectively.}
For video at 720p, MLO-STR achieves a staggering order-of-magnitude reduction in the \nnp delay compared to SLO, keeping it well below $10$ ms. The delay with MLO-NSTR is also consistently below that of SLO, although significantly underperforming MLO-STR. 
A resolution of 1080p can only be supported through MLO, with MLO-STR still guaranteeing a delay below $10$ ms.
Similar results were obtained with Poisson traffic, confirming the ability of MLO-STR to reduce the latency when the two links are symmetrically occupied regardless the traffic characteristics.

Figure \ref{st1} shows the delay performance for the asymmetric scenario considered in Figures \ref{delay3n} and \ref{nine}, i.e., that with primary and secondary link occupancy of $10\%$ and $70\%$, respectively. 
The experimental results obtained using Stadia's traffic at 720p exhibit similar qualitative trends as those employing Poisson traffic. Indeed, MLO-STR increases the delay over SLO by 85.5\%, a performance degradation even worse than the 68.6\% delay increase observed under Poisson traffic.

\textit{Findings:} Experiments using real application traces coupled with real channel occupancy traces confirmed our main findings originally obtained under Poisson traffic, namely:
\emph{(i)} MLO  achieves significant delay reduction over SLO, and may enable new applications whose traffic load cannot otherwise be delivered in a timely manner; 
\emph{(ii)} MLO-STR can suboptimally allocate packets to a secondary interface that is busier than the primary,  occasionally yielding even higher delays than SLO; For MLO-STR, such an effect can be further exacerbated under the real-world traffic considered due to batch packet arrivals; and
\emph{(iii)} By deferring the decision on which interface to allocate a packet to, MLO-STR+ yields the lowest delay in all scenarios considered.

\section{Channel Bonding}\label{bonding}

In previous sections, MLO was  using twice the  bandwidth of SLO, which certainly contributes to the observed gains. In this section, we use channel bonding to equalize the amount of spectrum bandwidth used by SLO and MLO, so as to better study the MLO performance gains stemming from contending over multiple narrow links vs. using a single wide one.

Channel bonding with preamble puncturing allows the use of multiple non contiguous 20 MHz channels. Each link uses an 80 MHz channel, containing four 20 MHz channels. From these, one is assigned as the primary, where backoff is performed, and all others are considered secondary. Prior to the backoff expiring, all secondary channels are checked, and all available channels are used for transmission (e.g., if three 20 MHz channels are open, transmission happens over 60 MHz). The same implementation of channel bonding with preamble puncturing used in \cite{barrachina2021wi} is considered, supporting also allocating multiple resource units to a single user as defined by IEEE 802.11be \cite{khorov2020current}.

We employ the same channels previously used, 36 and 100, as the primary channels of each link, and the subsequent channels as the secondary channels. For SLO, we use 40 MHz links (channels 36 and 40), and 80 MHz (36, 40, 44, and 48) links, while for MLO, we use 20 MHz (36 and 100) and 40 MHz (36-40 and 100-104) links, respectively. We use the same traffic loads as in the previous sections.

\subsection{Same bandwidth for SLO and MLO}

Previously, we used two 20 MHz links for MLO, thus doubling the bandwidth of the SLO link. In this section we study if SLO can achieve similar gains as MLO by just doubling its channel bandwidth, and thus both schemes use the same amount of spectrum.

We focus on the asymmetric link case only, under the same conditions as in previous sections. To generate the wider links, we consider the adjacent channels from the same dataset sample used in previous experiments so as to keep the existing temporal correlation between them.%

\begin{figure*}[ht]
\centering
\begin{subfigure}[b]{0.325\textwidth}
    \includegraphics[width = \textwidth]{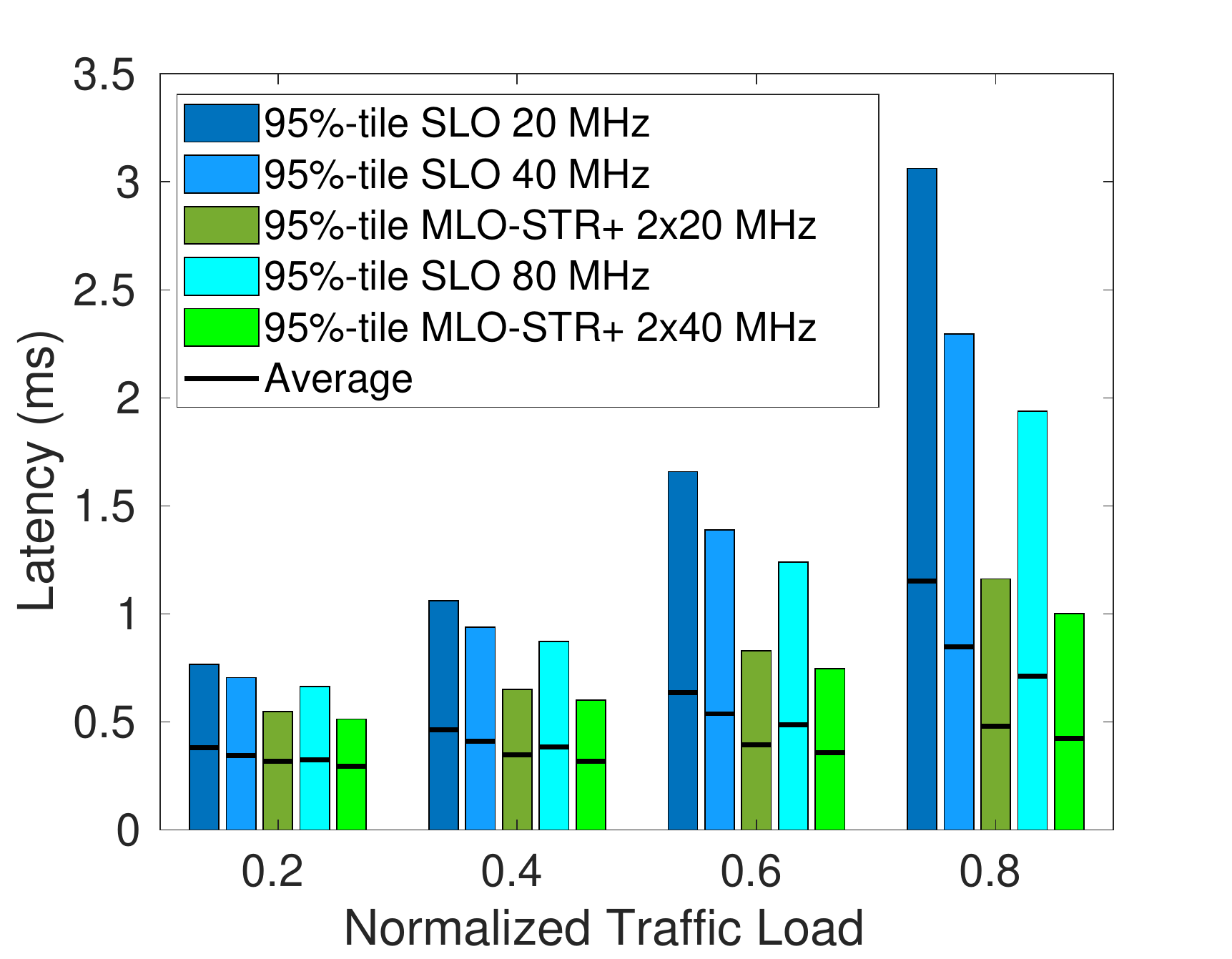}
    \caption{ Link occupancy of \{10\%, 40\%\}   }
    \label{delbw1}
\end{subfigure}
\begin{subfigure}[b]{0.325\textwidth}
    \includegraphics[width = \textwidth]{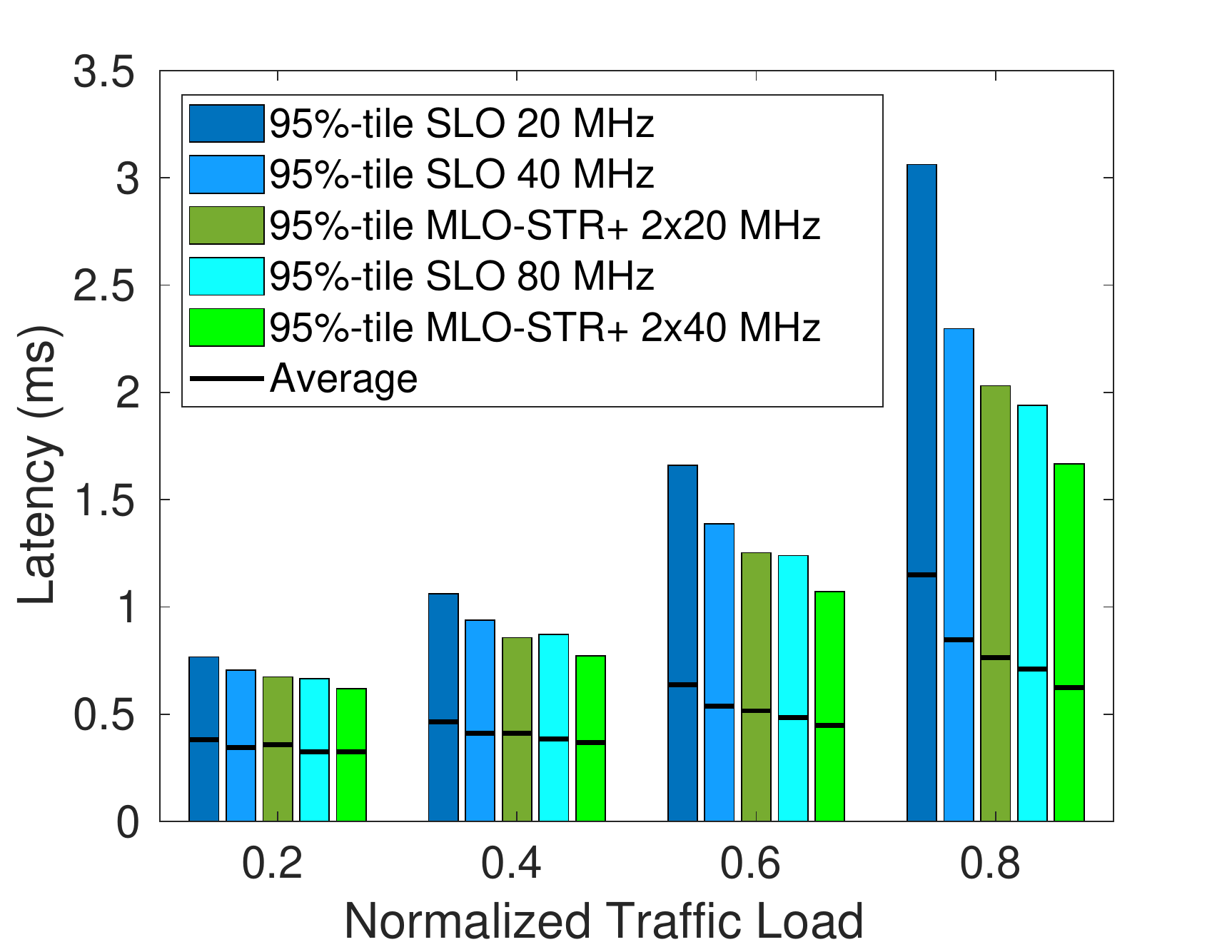}
    \caption{ Link occupancy of \{10\%, 70\%\}}
    \label{delbw2}
\end{subfigure}
\begin{subfigure}[b]{0.325\textwidth}
    \includegraphics[width = \textwidth]{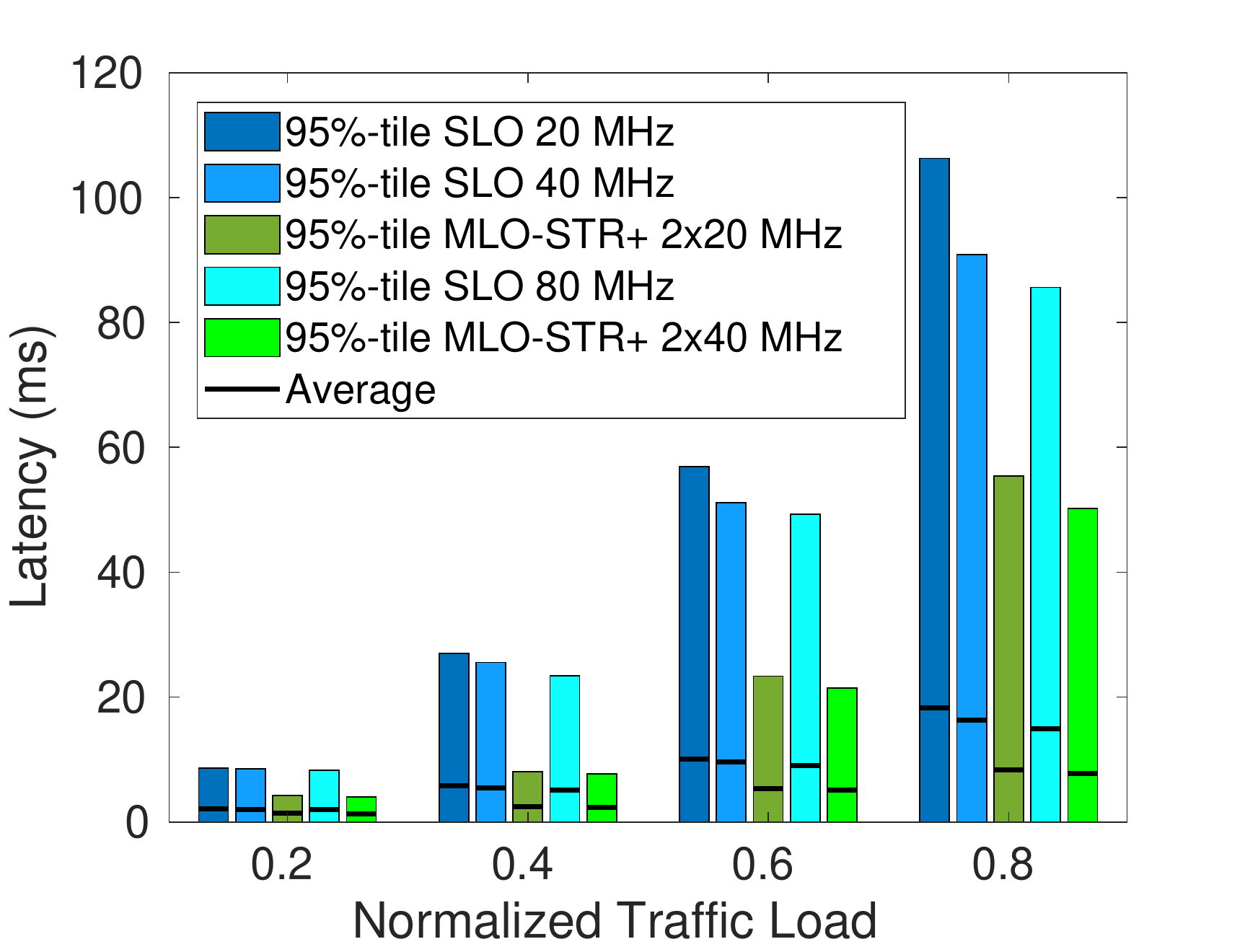}
    \caption{ Link occupancy of \{40\%, 70\%\}}
    \label{delbw3}
\end{subfigure}
\caption{Latency for asymmetric links vs. variable normalized traffic load under different channel bandwidth.}
\label{delbw}
\end{figure*}

Figure \ref{delbw} shows the average and \nnp delay for an increasing traffic load with different link bandwidths. As expected, using a link bandwidth of only 20 MHz leads to the highest delays for all schemes. Then, increasing the link bandwidth to 40 MHz leads to a delay decrease of up to 25\% in SLO, and 17.8\% in MLO-STR+. Comparing SLO performance at 40 MHz with MLO-STR+ using 2 links of 20 MHz each, we can observe that MLO-STR+ continues to outperform SLO. The same is observed for SLO 80 MHz and MLO-STR+ using 2 links of 40 MHz each. Moreover, we can observe that in Figures \ref{delbw1} and \ref{delbw3} MLO-STR+ with 20 MHz links achieves lower delays than SLO using a 80 MHz channel, i.e., MLO-STR+ improves SLO performance using half of the SLO bandwidth in total when the occupancy of the two links does not differ excessively. Otherwise, as shown in Figure \ref{delbw2}, this does not hold in the 10\%-70\% case since the secondary link is too busy, requiring to use the same bandwidth, i.e., two links of 40 MHz each, to outperform SLO.

\textit{Finding:} The use of higher bandwidth links leads to a decrease in the delay for both SLO and MLO. However, using two links of lower bandwidth, each with its own backoff, still leads to better results than a single link with higher bandwidth, showing that MLO-STR+ achieves its performance mainly due to running multiple backoff counters instead of one.

\subsection{Primary Channel Selection}

Channel bonding performance can vary depending on the primary channel used \cite{barrachina2021wi}. By allowing SLO to use wider channels than MLO to keep the same total bandwidth, SLO has more opportunities to find an emptier primary channel that could lead to lower delays than MLO.

We use the same setup as in the previous section, considering 40 and 80 MHz links, but each link is now free to select the less occupied 20 MHz channel as its primary channel in every spectrum sample.

\begin{figure}
    \centering
    \includegraphics[width=0.45\textwidth]{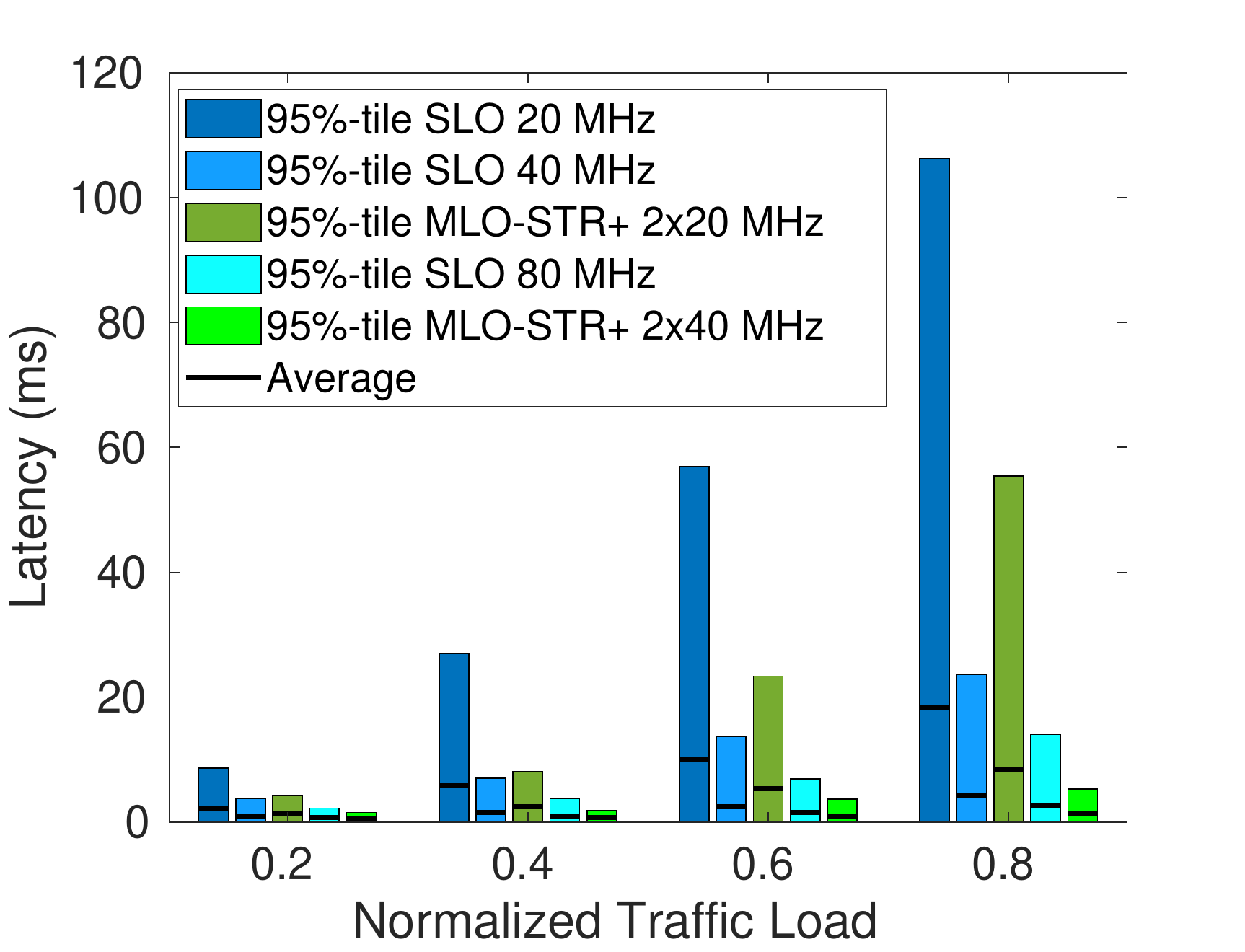}
    \caption{Latency for occupancy of \{40\%, 70\%\} with dynamic primary channel selection}
    \label{channeldynamic}
\end{figure}

Fig. \ref{channeldynamic} shows the delay for different channel widths when the less occupied channel of each link is selected as primary channel. Comparing the results with Fig. \ref{delbw3}, where the primary channel was fixed for each link, we can observe a significant delay reduction. For SLO, the delay for 40 and 80 MHz is reduced by a factor 3.8 and 6.1, respectively, compared to the case where the primary channel is fixed. Similarly, for MLO, for a 40 MHz link, the delay is reduced by a factor 9.5. Overall, a significant gain in delay is observed for both SLO and MLO-STR+ in Fig. \ref{channeldynamic} by allowing dynamic primary channel selection. It is specially remarkable that SLO 40 MHz results in lower latency than MLO-STR 2x20 MHz. The reason is that SLO is able to leverage the existence of channel 40, with an occupancy lower than the one of channel 36.

\textit{Findings:} Channel bonding and MLO operation both require careful selection of the channels used. A dynamic choice in the primary  channel shows up to 6.1$\times$ and 9.5$\times$ lower delays in SLO (80 MHz) and MLO-STR+ (2x40 MHz) respectively. Remarkably, there are cases where SLO can outperform MLO-STR+ when the extra available choices to set the primary channel allow it to find a much emptier channel than the ones used by MLO-STR+. Moreover, and focusing only on MLO, allowing to dynamically select the best primary channel in each link multiplies the latency improvements obtained by contending over multiple independent links, yielding almost an order of magnitude decrease compared to a static choice.

\subsection{Channel bonding and MLO-NSTR}

We now focus on MLO-NSTR, and study its performance using channel bonding with primary channel selection in the same conditions as MLO-STR+ in previous section. As MLO-NSTR's behavior is very similar to channel bonding (both using primary and secondary hierarchies), we would expect NSTR and SLO behavior to be similar if the bandwidths used are equal.

\begin{figure}
\centering
    \includegraphics[width = 0.45\textwidth]{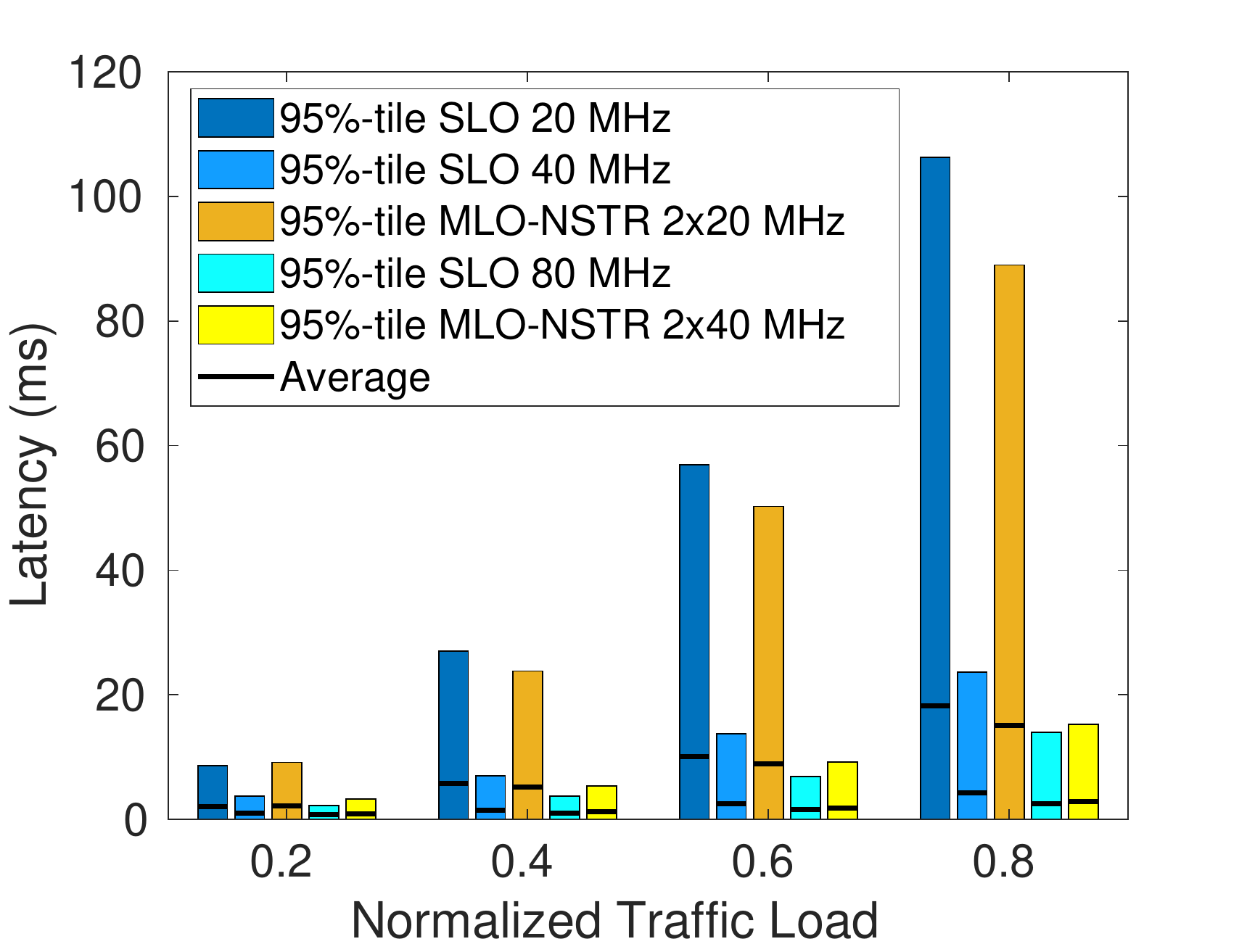}
    \caption{Latency for occupancy of \{40\%, 70\%\} with dynamic primary channel selection}
    \label{delbw22}

\end{figure}

Figure \ref{delbw22} shows the MLO-NSTR delay when the lowest occupancy primary 20 MHz channel is selected for each link. Much like Figure \ref{channeldynamic} for MLO-STR+, MLO-NSTR can greatly benefit from channel bonding, with its delay being reduced by a factor of up to 5.8$\times$ when switching from 20 MHz to 40 MHz links. However, SLO with a bandwidth of 80 MHz results in lower delays than 2 links of 40 MHz with MLO-NSTR, while MLO-STR+ still outperforms SLO. The reason is that SLO with an 80 MHz bandwidth uses 4 channels of 40\% occupancy, while MLO-NSTR has a secondary link using channels with 70\% occupancy, limiting the amount of times MLO-NSTR can transmit simultaneously, and the total effective bandwidth used.

\textit{Findings:} The use of a single backoff counter in its primary link prevents MLO-NSTR to benefit from parallel transmissions on both links, thus achieving a similar delay reduction as SLO in the best case.

\section{Related Work}\label{RW}

\subsubsection*{Multiple Radios}
The use of multiple radios or links has been studied for a variety of technologies and protocols. 5G Multi-Connectivity of a single device to multiple other technologies such as LTE and Wi-Fi has been investigated as a way to achieve higher capacity and meet ultra-reliability constraints \cite{odarchenko2018multilink,chandrashekar20165g,himayat2014multi}. 
In \cite{suer2020impact}, the importance of link homogeneity is studied, showing that using links that are too far apart in terms of latency will result in no gains over single link communication.
Multi-Path TCP has also been considered as a way to enable the use of TCP over multiple connections of Wi-Fi, 3G and LTE \cite{deng2014wifi,nguyen2014cross,paasch2012exploring,pokhrel2018improving}. In \cite{amend2019framework}, a framework for Multi-Path with multi-connectivity across different networks is presented, which also shows a decrease in bitrate when links have differences in their latencies. In \cite{viernickel2018multipath}, a Multi-Path implementation for QUIC is presented, also discussing the impact of heterogeneous links, and how adding high latency secondary paths can lead to worse performance. 

The use of  multiple radios focusing only on Wi-Fi deployments has been studied as well, seeking to improve reliability \cite{schwarzenberg2018quantifying} and reduce  latency \cite{kondo2020low} through cooperative links. Handoff delays can also be reduced by dedicating one radio to data exchange while another one serves management frames \cite{jin2013seamless, ramachandran2006make, brik2005eliminating}. Insights learned through such works have provided the basis for the current Wi-Fi standardization efforts to support multiple radios and Multi-link Operation in IEEE 802.11be. 

\subsubsection*{Multi-Link Operation}
Multi-Link Operation ---as the key feature of IEEE 802.11be--- has already received significant attention from the Wi-Fi community. The feasibility of the Simultaneous Transmission and Reception mode depending on the amount of cross-link interference is studied in \cite{levitsky2020study}, assessing the minimum spectral distance required between links, and showing that 100 MHz is enough to ensure proper packet reception, validating the 200 MHz separation between channels considered in our work. The  evaluation of the impact that multi-link transmission has on the worst case latency for real time applications is shown in \cite{naik2021can}, showing that only using two links already leads to an order of magnitude delay reduction in the 90th percentile delay in some cases, which is confirmed by our results. STR delay for industrial and latency-bound settings is also investigated in \cite{lacalle2021analysis}, showing that a secondary link leads to halving the average delay and almost halving the worst case delay. 

\subsubsection*{Coexistence and Traffic Differentiation}
The interplay between SLO and MLO devices and its impact on latency is studied in \cite{carrascosa2022performance} and \cite{adhikari2022analysis}. NSTR coexistence with legacy devices is also studied in \cite{murti2022multilink,korolev2022study}. The interplay between multiple STR nodes is studied in \cite{carrascosa2022understanding} Different implementations of Multi-Link Operation are studied in \cite{song2021performance,chen2022overview, yang2019ap}, analysing the impact that each of them has on the WLAN throughput, the latter finding a throughput increase of 200\% and 80\% for STR and NSTR over SLO, respectively, which aligns with our findings of MLO-STR with symmetrical occupancy, and MLO-NSTR for a primary of 10\% occupancy . 
Finally, traffic allocation policies using multiple links are considered in \cite{lopez2021ieee}, and an adaptation of EDCA with MLO NSTR devices can be found in \cite{park2021latency} and in \cite{9967996}. Real time application traffic is also discussed in \cite{bankov2021use}, where different frequency resource allocation schemes for MLO are proposed. 

\subsubsection*{Performance Analysis}
Most of the cited references have studied MLO performance through simulations, although there are also some cases in which new analytical models are derived to study MLO performance. With the exception of \cite{bellalta2022analysis} where the focus is placed on the delay analysis under finite load conditions, all other papers \cite{song2021performance, yang2019ap, korolev2022analytical, chen2022overview, 9967996} focus on the MLO efficiency under saturation (full-buffer) throughput conditions.

To the best of our knowledge, this paper is the first work studying the performance of Multi-Link Operation using real spectrum occupancy measurements. While our results confirm the potential latency gains of MLO seen in  the state of the art, the use of real spectrum measurements offers new and unique insights on MLO performance otherwise not possible.

\section{Conclusions}\label{conc}

To the best of our knowledge, this paper is the first experimental study of throughput and latency for MLO. Using a dataset containing real-world channel occupancy measurements in the 5 GHz spectrum, we studied throughput and latency performance of two MLO channel access modes, MLO-STR and MLO-NSTR.

We demonstrated that MLO can enable new latency-sensitive applications, whose traffic load cannot otherwise be delivered in a timely manner through SLO. We showed that when both links are similarly occupied, both MLO modes can reduce the \nnp latency by nearly one order of magnitude. In contrast, with asymmetrically occupied links, we surprisingly found that MLO-STR, the mode with superior throughput performance, can sometimes yield higher worst-case latency than SLO. We proposed a \emph{deferred decision} enhancement to MLO-STR that overcomes this limitation.  We studied performance on traffic from a real delay-sensitive gaming application to couple real channel experiments with real application experiments. Finally, we showed how the gains attained by MLO can further grow when using channel bonding.

An important extension of this work would attempt at capturing the behavior of various MLO modes analytically, rather than via simulations, thereby allowing a more generalized comparison and wide reproducibility of the results. We refer the reader to \cite{bellalta2022analysis} for a first attempt at the latter, and to \cite{chen2022overview,carrascosa2022understanding} for fresh summaries of the MLO standardization process. With Wi-Fi 7 defined and MLO up and running, beyond-802.11be technologies are expected to operate in new frequency bands \cite{ResCor22} and/or augment the spatial reuse of the old ones through advanced AP coordination \cite{garcia2021ieee}. Looking ahead, any new features being introduced in Wi-Fi 8\cite{giordano2023will} should be conceived atop MLO, and their performance studied when paired with the latter.

\bibliographystyle{IEEEtran}

\bibliography{main.bbl}

\begin{thebibliography}{10}
\providecommand{\url}[1]{#1}
\csname url@samestyle\endcsname
\providecommand{\newblock}{\relax}
\providecommand{\bibinfo}[2]{#2}
\providecommand{\BIBentrySTDinterwordspacing}{\spaceskip=0pt\relax}
\providecommand{\BIBentryALTinterwordstretchfactor}{4}
\providecommand{\BIBentryALTinterwordspacing}{\spaceskip=\fontdimen2\font plus
\BIBentryALTinterwordstretchfactor\fontdimen3\font minus
  \fontdimen4\font\relax}
\providecommand{\BIBforeignlanguage}[2]{{%
\expandafter\ifx\csname l@#1\endcsname\relax
\typeout{** WARNING: IEEEtran.bst: No hyphenation pattern has been}%
\typeout{** loaded for the language `#1'. Using the pattern for}%
\typeout{** the default language instead.}%
\else
\language=\csname l@#1\endcsname
\fi
#2}}
\providecommand{\BIBdecl}{\relax}
\BIBdecl

\bibitem{carrascosa2021experimentalPublished}
M.~Carrascosa, G.~Geraci, E.~Knightly, and B.~Bellalta, ``{An Experimental
  Study of Latency for IEEE 802.11 be Multi-link Operation},'' in \emph{{IEEE
  ICC 2022- IEEE International Conference on Communications}}.\hskip 1em plus
  0.5em minus 0.4em\relax {IEEE}, {2022}, pp. {2507--2512}.

\bibitem{adame2021time}
T.~Adame, M.~Carrascosa-Zamacois, and B.~Bellalta, ``{Time-sensitive networking
  in IEEE 802.11 be: On the way to low-latency WiFi 7},'' \emph{{Sensors}},
  vol.~{21}, no.~{15}, p. {4954}, {2021}.

\bibitem{pei2016wifi}
C.~Pei, Y.~Zhao, G.~Chen, R.~Tang, Y.~Meng, M.~Ma, K.~Ling, and D.~Pei, ``{WiFi
  can be the weakest link of round trip network latency in the wild},'' in
  \emph{{IEEE INFOCOM 2016-The 35th Annual IEEE International Conference on
  Computer Communications}}.\hskip 1em plus 0.5em minus 0.4em\relax {IEEE},
  {2016}, pp. {1--9}.

\bibitem{draft11be}
``{IEEE P802.11be/D1.0 Draft Standard for Information technology—
  Telecommunications and information exchange between systems Local and
  metropolitan area networks— Specific requirements. Part 11: Wireless LAN
  Medium Access Control (MAC) and Physical Layer (PHY) Specifications.
  Amendment 8: Enhancements for extremely high throughput (EHT)},'' {May}
  {2021}.

\bibitem{lopez2022multi}
A.~L{\'o}pez-Ravent{\'o}s and B.~Bellalta, ``{Multi-link Operation in IEEE
  802.11 be WLANs},'' \emph{IEEE Wireless Communications}, 2022.

\bibitem{garcia2021ieee}
A.~Garcia-Rodriguez, D.~Lopez-Perez, L.~Galati-Giordano, and G.~Geraci, ``{IEEE
  802.11 be: Wi-Fi 7 Strikes Back},'' \emph{{IEEE Communications Magazine}},
  vol.~{59}, no.~{4}, pp. {102--108}, {2021}.

\bibitem{khorov2020current}
E.~Khorov, I.~Levitsky, and I.~F. Akyildiz, ``{Current status and directions of
  IEEE 802.11 be, the future Wi-Fi 7},'' \emph{{IEEE Access}}, vol.~{8}, pp.
  {88\,664--88\,688}, {2020}.

\bibitem{hoefel2020ieee}
R.~P.~F. Hoefel, ``{IEEE 802.11 be: Throughput and Reliability Enhancements for
  Next Generation WI-FI Networks},'' in \emph{{2020 IEEE 31st Annual
  International Symposium on Personal, Indoor and Mobile Radio
  Communications}}.\hskip 1em plus 0.5em minus 0.4em\relax {IEEE}, {2020}, pp.
  {1--7}.

\bibitem{yang2020survey}
M.~Yang and B.~Li, ``{Survey and perspective on extremely high throughput (EHT)
  WLAN—IEEE 802.11 be},'' \emph{{Mobile Networks and Applications}},
  vol.~{25}, no.~{5}, pp. {1765--1780}, {2020}.

\bibitem{lopez2019ieee}
D.~L{{\'o}}pez-P{{\'e}}rez, A.~Garcia-Rodriguez, L.~Galati-Giordano,
  M.~Kasslin, and K.~Doppler, ``{IEEE 802.11 be extremely high throughput: The
  next generation of Wi-Fi technology beyond 802.11 ax},'' \emph{{IEEE
  Communications Magazine}}, vol.~{57}, no.~{9}, pp. {113--119}, {2019}.

\bibitem{deng2020ieee}
C.~Deng, X.~Fang, X.~Han, X.~Wang, L.~Yan, R.~He, Y.~Long, and Y.~Guo, ``{IEEE
  802.11 be Wi-Fi 7: New challenges and opportunities},'' \emph{{IEEE
  Communications Surveys \& Tutorials}}, vol.~{22}, no.~{4}, pp. {2136--2166},
  {2020}.

\bibitem{song2021performance}
T.~Song and T.~Kim, ``{Performance Analysis of Synchronous Multi-Radio
  Multi-Link MAC Protocols in IEEE 802.11 be Extremely High Throughput
  WLANs},'' \emph{{Applied Sciences}}, vol.~{11}, no.~{1}, p. {317}, {2021}.

\bibitem{yang2019ap}
M.~Yang, B.~Li, Z.~Yan, and Y.~Yan, ``{AP Coordination and Full-duplex enabled
  Multi-band Operation for the Next Generation WLAN: IEEE 802.11 be (EHT)},''
  in \emph{{2019 11th International Conference on Wireless Communications and
  Signal Processing (WCSP)}}.\hskip 1em plus 0.5em minus 0.4em\relax {IEEE},
  {2019}, pp. {1--7}.

\bibitem{levitsky2020study}
I.~Levitsky, Y.~Okatev, and E.~Khorov, ``{Study on Simultaneous Transmission
  and Reception on Multiple Links in IEEE 802.11 be networks},'' in \emph{{2020
  International Conference Engineering and Telecommunication (En\&T)}}.\hskip
  1em plus 0.5em minus 0.4em\relax {IEEE}, {2020}, pp. {1--4}.

\bibitem{lopez2021ieee}
{\'A}.~L{\'o}pez-Ravent{\'o}s and B.~Bellalta, ``{IEEE 802.11 be multi-link
  operation: When the best could be to use only a single interface},'' in
  \emph{2021 19th Mediterranean Communication and Computer Networking
  Conference (MedComNet)}.\hskip 1em plus 0.5em minus 0.4em\relax IEEE, 2021,
  pp. 1--7.

\bibitem{park2021latency}
H.~Park and C.~You, ``{Latency Impact for Massive Real-Time Applications on
  Multi Link Operation},'' in \emph{{2021 IEEE Region 10 Symposium
  (TENSYMP)}}.\hskip 1em plus 0.5em minus 0.4em\relax {IEEE}, {2021}, pp.
  {1--5}.

\bibitem{bankov2021use}
D.~Bankov, A.~Lyakhov, E.~Khorov, and K.~Chemrov, ``{On the Use of Multilink
  Access Methods to Support Real-Time Applications in Wi-Fi Networks},''
  \emph{{Journal of Communications Technology and Electronics}}, vol.~{66},
  no.~{12}, pp. {1476--1484}, {2021}.

\bibitem{naik2021can}
G.~Naik, D.~Ogbe, and J.-M.~J. Park, ``{Can Wi-Fi 7 support real-time
  applications? On the impact of multi link aggregation on latency},'' in
  \emph{{ICC 2021-IEEE International Conference on Communications}}.\hskip 1em
  plus 0.5em minus 0.4em\relax {IEEE}, {2021}, pp. {1--6}.

\bibitem{lacalle2021analysis}
G.~Lacalle, I.~Val, O.~Seijo, M.~Mendicute, D.~Cavalcanti, and
  J.~Perez-Ramirez, ``{Analysis of latency and reliability improvement with
  multi-link operation over 802.11},'' in \emph{{2021 IEEE 19th International
  Conference on Industrial Informatics (INDIN)}}.\hskip 1em plus 0.5em minus
  0.4em\relax {IEEE}, {2021}, pp. {1--7}.

\bibitem{schwarzenberg2018quantifying}
N.~Schwarzenberg, A.~Wolf, N.~Franchi, and G.~Fettweis, ``{Quantifying the gain
  of multi-connectivity in wireless LAN},'' in \emph{{2018 European Conference
  on Networks and Communications (EuCNC)}}.\hskip 1em plus 0.5em minus
  0.4em\relax {IEEE}, {2018}, pp. {16--20}.

\bibitem{kondo2020low}
Y.~Kondo, Y.~Hiroyuki, and H.~Yokoyama, ``{A Low Latency Transmission Control
  for Multi-link WLAN},'' in \emph{{2020 29th International Conference on
  Computer Communications and Networks (ICCCN)}}.\hskip 1em plus 0.5em minus
  0.4em\relax {IEEE}, {2020}, pp. {1--6}.

\bibitem{barrachina2020wi}
S.~Barrachina-Mu{{\~n}}oz, B.~Bellalta, and E.~Knightly, ``{Wi-Fi All-Channel
  Analyzer},'' in \emph{{Proceedings of the 14th International Workshop on
  Wireless Network Testbeds, Experimental evaluation \& Characterization}},
  {2020}, pp. {72--79}.

\bibitem{barrachina2021wi}
S.~Barrachina-Mu{{\~n}}oz, B.~Bellalta, and E.~W. Knightly, ``{Wi-fi channel
  bonding: An all-channel system and experimental study from urban hotspots to
  a sold-out stadium},'' \emph{{IEEE/ACM Transactions on Networking}},
  vol.~{29}, no.~{5}, pp. {2101--2114}, {2021}.

\bibitem{carrascosa2022cloud}
M.~Carrascosa and B.~Bellalta, ``{Cloud-gaming: Analysis of google stadia
  traffic},'' \emph{{Computer Communications}}, vol. 188, pp. 99--116, 2022.

\bibitem{bellalta2020low}
B.~Bellalta, ``{On the Low-latency Region of Best-effort Links for
  Delay-Sensitive Streaming Traffic},'' \emph{IEEE Communications Letters},
  vol.~25, no.~3, pp. 970--974, 2020.

\bibitem{bellalta2022analysis}
B.~Bellalta, M.~Carrascosa, L.~Galati-Giordano, and G.~Geraci, ``{Delay
  Analysis of {IEEE 802.11be}} multi-link operation under finite load,''
  \emph{IEEE Wireless Communications Letters}, 2023.

\bibitem{odarchenko2018multilink}
R.~Odarchenko, R.~Aguiar, B.~Altman, and Y.~Sulema, ``{Multilink approach for
  the content delivery in 5G networks},'' in \emph{{2018 International
  Scientific-Practical Conference Problems of Infocommunications. Science and
  Technology (PIC S\&T)}}.\hskip 1em plus 0.5em minus 0.4em\relax {IEEE},
  {2018}, pp. {140--144}.

\bibitem{chandrashekar20165g}
S.~Chandrashekar, A.~Maeder, C.~Sartori, T.~H{{\"o}}hne, B.~Vejlgaard, and
  D.~Chandramouli, ``{5G multi-RAT multi-connectivity architecture},'' in
  \emph{{2016 IEEE International Conference on Communications Workshops
  (ICC)}}.\hskip 1em plus 0.5em minus 0.4em\relax {IEEE}, {2016}, pp.
  {180--186}.

\bibitem{himayat2014multi}
N.~Himayat, S.-p. Yeh, A.~Y. Panah, S.~Talwar, M.~Gerasimenko, S.~Andreev, and
  Y.~Koucheryavy, ``{Multi-radio heterogeneous networks: Architectures and
  performance},'' in \emph{{2014 International Conference on Computing,
  Networking and Communications (ICNC)}}.\hskip 1em plus 0.5em minus
  0.4em\relax {IEEE}, {2014}, pp. {252--258}.

\bibitem{suer2020impact}
M.-T. Suer, C.~Thein, H.~Tchouankem, and L.~Wolf, ``{Impact of link
  heterogeneity and link correlation on Multi-Connectivity scheduling schemes
  for reliable Low-Latency communication},'' in \emph{{2020 IEEE International
  Conference on Communications Workshops (ICC Workshops)}}.\hskip 1em plus
  0.5em minus 0.4em\relax IEEE, 2020, pp. 1--6.

\bibitem{deng2014wifi}
S.~Deng, R.~Netravali, A.~Sivaraman, and H.~Balakrishnan, ``{WiFi, LTE, or
  both? Measuring multi-homed wireless internet performance},'' in
  \emph{{Proceedings of the 2014 Conference on Internet Measurement
  Conference}}, {2014}, pp. {181--194}.

\bibitem{nguyen2014cross}
K.~Nguyen, Y.~Ji, and S.~Yamada, ``{A cross-layer approach for improving WiFi
  performance},'' in \emph{{2014 International Wireless Communications and
  Mobile Computing Conference (IWCMC)}}.\hskip 1em plus 0.5em minus 0.4em\relax
  {IEEE}, {2014}, pp. {458--463}.

\bibitem{paasch2012exploring}
C.~Paasch, G.~Detal, F.~Duchene, C.~Raiciu, and O.~Bonaventure, ``{Exploring
  mobile/WiFi handover with multipath TCP},'' in \emph{{Proceedings of the 2012
  ACM SIGCOMM workshop on Cellular networks: operations, challenges, and future
  design}}, {2012}, pp. {31--36}.

\bibitem{pokhrel2018improving}
S.~R. Pokhrel and M.~Mandjes, ``{Improving multipath TCP performance over WiFi
  and cellular networks: An analytical approach},'' \emph{{IEEE Transactions on
  Mobile Computing}}, vol.~{18}, no.~{11}, pp. {2562--2576}, {2018}.

\bibitem{amend2019framework}
M.~Amend, E.~Bogenfeld, M.~Cvjetkovic, V.~Rakocevic, M.~Pieska, A.~Kassler, and
  A.~Brunstrom, ``{A framework for multiaccess support for unreliable internet
  traffic using multipath dccp},'' in \emph{{2019 IEEE 44th Conference on Local
  Computer Networks (LCN)}}.\hskip 1em plus 0.5em minus 0.4em\relax IEEE, 2019,
  pp. 316--323.

\bibitem{viernickel2018multipath}
T.~Viernickel, A.~Froemmgen, A.~Rizk, B.~Koldehofe, and R.~Steinmetz,
  ``{Multipath QUIC: A deployable multipath transport protocol},'' in
  \emph{{2018 IEEE International Conference on Communications (ICC)}}.\hskip
  1em plus 0.5em minus 0.4em\relax IEEE, 2018, pp. 1--7.

\bibitem{jin2013seamless}
S.~Jin and S.~Choi, ``{A seamless handoff with multiple radios in IEEE 802.11
  WLANs},'' \emph{{IEEE Transactions on Vehicular Technology}}, vol.~{63},
  no.~{3}, pp. {1408--1418}, {2013}.

\bibitem{ramachandran2006make}
K.~Ramachandran, S.~Rangarajan, and J.~C. Lin, ``{Make-before-break mac layer
  handoff in 802.11 wireless networks},'' in \emph{{2006 IEEE International
  Conference on Communications}}, vol.~{10}.\hskip 1em plus 0.5em minus
  0.4em\relax {IEEE}, {2006}, pp. {4818--4823}.

\bibitem{brik2005eliminating}
V.~Brik, A.~Mishra, and S.~Banerjee, ``{Eliminating handoff latencies in 802.11
  WLANs using multiple radios: Applications, experience, and evaluation},'' in
  \emph{{Proceedings of the 5th ACM SIGCOMM conference on Internet
  Measurement}}, {2005}, pp. {27--27}.

\bibitem{carrascosa2022performance}
M.~Carrascosa-Zamacois, L.~Galati-Giordano, A.~Jonsson, G.~Geraci, and
  B.~Bellalta, ``{Performance and Coexistence Evaluation of IEEE 802.11 be
  Multi-link Operation},'' in \emph{{Proceedings of the 2023 IEEE Wireless
  Communications and Networking Conference }}, {2023}, pp. {1--6}.

\bibitem{adhikari2022analysis}
S.~Adhikari and S.~Verma, ``{Analysis of Multilink in IEEE 802.11 be},''
  \emph{{IEEE Communications Standards Magazine}}, vol.~6, no.~3, pp. 52--58,
  2022.

\bibitem{murti2022multilink}
W.~Murti and J.-H. Yun, ``{Multilink Operation in IEEE 802.11 be Wireless LANs:
  Backoff Overflow Problem and Solutions},'' \emph{{Sensors}}, vol.~22, no.~9,
  p. 3501, 2022.

\bibitem{korolev2022study}
N.~Korolev, I.~Levitsky, I.~Startsev, B.~Bellalta, and E.~Khorov, ``{Study of
  Multi-link Channel Access without Simultaneous Transmit and Receive in IEEE
  802.11 be Networks},'' \emph{IEEE Access}, 2022.

\bibitem{carrascosa2022understanding}
M.~Carrascosa-Zamacois, G.~Geraci, L.~Galati-Giordano, A.~Jonsson, and
  B.~Bellalta, ``{Understanding Multi-link Operation in Wi-Fi 7: Performance,
  Anomalies, and Solutions},'' \emph{arXiv preprint arXiv:2210.07695}, 2022.

\bibitem{chen2022overview}
C.~Chen, X.~Chen, D.~Das, D.~Akhmetov, and C.~Cordeiro, ``{Overview and
  Performance Evaluation of Wi-Fi 7},'' \emph{{IEEE Communications Standards
  Magazine}}, vol.~6, no.~2, pp. 12--18, 2022.

\bibitem{9967996}
N.~Korolev, I.~Levitsky, I.~Startsev, B.~Bellalta, and E.~Khorov, ``{Study of
  Multi-link Channel Access without Simultaneous Transmit and Receive in IEEE
  802.11be Networks},'' \emph{IEEE Access}, pp. 1--1, 2022.

\bibitem{korolev2022analytical}
N.~Korolev, I.~Levitsky, and E.~Khorov, ``{Analytical Model of Multi-link
  Operation in Saturated Heterogeneous Wi-Fi 7 Networks},'' \emph{{IEEE
  Wireless Communications Letters}}, 2022.

\bibitem{ResCor22}
E.~Reshef and C.~Cordeiro, ``{Future Directions for {Wi-Fi 8} and Beyond},''
  \emph{IEEE Communications Magazine}, pp. 1--7, 2022.

\bibitem{giordano2023will}
L.~Galati-Giordano, G.~Geraci, M.~Carrascosa, and B.~Bellalta, ``{What Will
  Wi-Fi 8 Be? A Primer on IEEE 802.11bn Ultra High Reliability},'' \emph{arXiv
  preprint arXiv:2303.10442}, 2023.

\end{thebibliography}

\end{document}